\documentclass[%
 aps,
 pra,
 showkeys,
 amsmath,amssymb,floatfix,
reprint,%
]{revtex4-1}
 
\usepackage{graphicx}
\usepackage{dcolumn}
\usepackage{bm}
\usepackage{color}

\graphicspath{{figures/}}

\usepackage{hyperref}
\usepackage{cancel}

\usepackage{multirow}
\hypersetup{
   colorlinks,
   linkcolor={blue},
   citecolor={blue},
   urlcolor={blue} 
}

\usepackage{comment}
\usepackage[dvipsnames]{xcolor}

\begin{document}

\title[
]{
Robust parity-mixed superconductivity in disordered monolayer transition metal dichalcogenides 
}

\author{David M\"{o}ckli
}
\email[E-mail me at: ]{d.mockli@gmail.com}
\affiliation{
The Racah Institute of Physics, The Hebrew University of Jerusalem, Jerusalem 9190401, Israel} 
\author{Maxim Khodas} 
\affiliation{
The Racah Institute of Physics, The Hebrew University of Jerusalem, Jerusalem 9190401, Israel}

\date{\today}

\begin{abstract}
Monolayer NbSe$_2$ is a nodal topological Ising superconductor at magnetic in-plane fields exceeding the Pauli limit, with nodal points strictly on high symmetry lines in the Brillouin zone. 
Here, we use a combined numerical and group-theoretical approach in real-space to characterize the unconventional superconducting state in monolayer transition metal dichalcogenides.
Even with a conventional pairing interaction, the superconducting state is intrinsically parity-mixed and robust against on-site disorder.  The interplay between the Zeeman magnetic field, strong spin-orbit interaction, and electronic orbital content confer the unique superconducting and topological properties. The discussion also extends to strongly hole-doped MoS$_2$ and its relatives. 
\end{abstract}
  



\maketitle

\section{Introduction}

Magnetic fields and impurities affect superconductivity in various ways.
For conventional Bardeen-Cooper-Schrieffer (BCS) superconductors, magnetic fields are detrimental to superconductivity in mainly two ways: the first is due to the coupling of the superconducting order parameter to the charge that confines the electrons to orbits leading to "orbital limiting"; and the second originates from the Zeeman coupling to the spin by breaking up Cooper pairs leading to "paramagnetic limiting". 
In most superconductors, the paramagnetic effect is negligible, because the orbital upper critical field $H_{c2}$ is much lower than the paramagnetic limit $H_\mathrm{P}$. 
However, for in-plane magnetic fields applied to quasi-2D materials, electronic dynamics is restricted to the basal plane such that the orbital effect is negligible, and the critical field is given by $H_\mathrm{P}$. 
Conversely, conventional BCS superconductors are robust against scalar disorder according to Anderson's criteria \cite{Anderson1959}, whereas disorder usually suppresses unconventional pairing states \cite{Force1963,Mackenzie2003,Balatsky2006}. 
In this paper, we obtain the unconventional superconducting state in monolayer NbSe$_2$ and transition metal dichalcogenides in general -- which withstand in-plane magnetic fields beyond the paramagnetic limit \cite{Khestanova2018,Ugeda2015,Bawden2016,Dvir2017,Xing2017,Sohn2018,Nakata2018} -- and investigate how the disorder affects the parity-mixed superconducting state.

Transition metal dichalcogenides (TMD) exhibit chemical versatility as compared to graphene \cite{Chhowalla2013}.
They are layered Van-der-Waals materials of chemical structure MX$_2$, where M is a group 4-10 transition metal and X is a group 16 chalcogen atom ($\mathrm{X}=\mathrm{S,\, Se}$ or Te).
We focus on the hexagonal monolayer polytype (1H) with crystal point group $D_{3h}$. Among the most famous examples is the group-6 direct band-gap semiconductor molybdenum disulfide (MoS$_2$) that has promising applications in next-generation electronic devices \cite{Radisavljevic2011,Wu2014,Lu2015}, and the superconducting metal niobium diselenide (NbSe$_2$) known for its wealth of electronic and magnetic phases  \cite{Calandra2009,Xi2015a,Ugeda2015,Flicker2015,Zheng2018}.  
The crystal of TMDs in bulk form possesses a global inversion center, but a monolayer is non-centrosymmetric. 
Viewing the monolayer crystal from above (along the $c$-axis), 1H-TMDs forms a hexagonal lattice similar to graphene, but with two inequivalent sublattices that break inversion \cite{Kormanyos2015}.
The lack of a definite parity allows for the emergence of unconventional superconducting states \cite{Smidman2017,Gorkov2001,Liu2016,Yip2014,Bauer2012,Kozii2015,Sigrist2014}, and the potential to realise topological superconductivity \cite{Yuan2014,Zhou2016,Fischer2018,Hsu2017,Xu2014,He2018}.     
Although parity lacks, the basal mirror symmetry restricts the crystal electric field to in-plane directions, such that the spin-orbit magnetic induction points in the out-of-plane direction. This peculiar form of spin-orbit coupling (SOC) locks the spins in the out-of-plane plane direction. For this reason, the superconducting state that develops from the normal state is frequently referred to as Ising superconductivity, or Zeeman protected superconductivity \cite{Ugeda2015,Zhou2016,Shimahara2000,DelaBarrera2018,Saito2016,Navarro-Moratalla2016,Xi2015}.  

\begin{figure*} 
\centering
\includegraphics[width=0.9\textwidth]{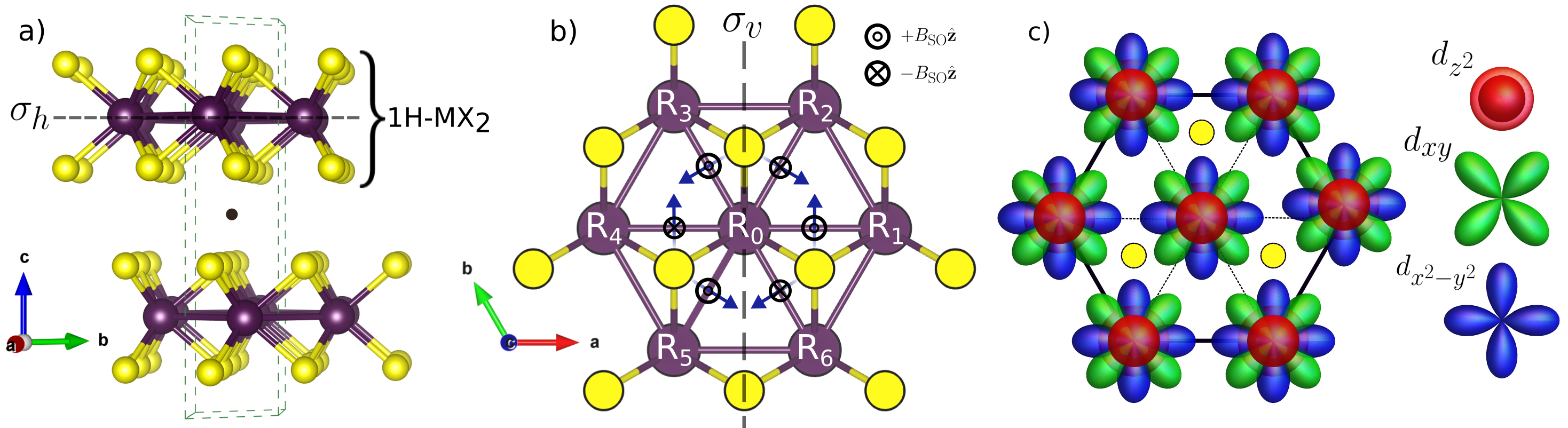}
\caption{\label{fig:crystal} Crystal structure of H-polytype TMDs with trigonal prismatic coordination. Violet atoms show the transition metal M and yellow atoms are the chalcogens X. a) The dashed green box shows the unit cell of 2H-MX$_2$. The black dot indicates the inversion center present in 2H-MX$_2$. b) Top view of 1H-MX$_2$. Inversion lacks, but there is basal mirror $\sigma_h$ plane and three perpendicular mirrors to the plane according to the point group $C_{3v}$. 
One of the $C_{3v}$ mirrors is indicated by $\sigma_v$. The blue arrows show the direction of the in-plane crystal field, and the symbols $\bigodot$  and $\otimes$ indicate the anti-symmetric out-of-plane Ising spin-orbit magnetic induction $B_\mathrm{SO}$.  
c) The minimal set of $4d$ orbitals used for the tight-binding model. The orbitals are $d_{z^2}$ (red), $d_{xy}$ (green), and $d_{x^2-y^2}$ (blue). Because of the basal mirror symmetry $\sigma_h$, the $4d_{xy}$ and $4d_{xz}$ orbitals do not participate. The use of Vesta software aided in the elaboration of figures (a) and (b) \cite{Momma2011}.}
\end{figure*}

To be specific, we take monolayer NbSe$_2$ as our basic model, which on the band structure level is qualitatively similar to heavily hole-doped monolayer MoS$_2$. Therefore, our analysis is relevant not only for metallic TMDs such as NbSe$_2$ and TaS$_2$, but extends to all the semiconducting cousins of MoS$_2$ in the strongly hole-doped ($p$-doped) regime.

Previous studies of the effect of scalar impurities on the superconducting state in TMD monolayers used a minimalist model mimicking graphene with two inequivalent sublattices,
and predict a suppression of the critical field by dilute impurities due to inter-valley scattering \cite{Sosenko2017,Ilic2017}. 
On the other hand, strong SOC together with a residual chiral symmetry is known to protect unconventional order parameters against disorder respecting this symmetry \cite{Michaeli2012}. Here we use a realistic three-orbital tight-binding model to investigate the structure of the superconducting state that emerges from the Ising spin-locked normal state Hamiltonian \cite{Bawden2016,Nakata2018}, and focus on the effect of a paramagnetic limiting in-plane magnetic field, and the role of scalar impurities taking into account the orbital degree of freedom. 
We employ a combined group-theoretical and numerical analysis of the symmetries of the emergent unconventional superconducting state. While group theory provides us with a classification of the allowed pairing symmetries, the self-consistent real-space Bogoliubov-deGennes simulations (BdG) finds the amplitude and the structure of the superconducting pairing correlations. 
We attribute the remarkable robustness of the unconventional superconducting state to the Ising SOC and the orbital wave-function orthogonality.

This paper is organized as follows: in section \ref{sec:tbmodel}, we introduce the normal state real-space tight-binding Hamiltonian of monolayer TMDs. 
In section \ref{sec:unconventional}, we introduce the basic elements of unconventional Ising superconductivity and provide a group-theoretical analysis of the on-site pairing correlations. 
In section \ref{sec:numerical}, we present the essentials of the Chebyshev-BdG expansion method (also known as the kernel polynomial method), which is used to solve the real-space Hamiltonian. In section \ref{sec:results}, we show the results of the numerical simulations, focusing on the superconducting state that self-consistently emerges from the normal state correlations, the effect of an in-plane paramagnetically limiting field and on-site scalar impurities.
In section \ref{sec:discussion}, we discuss the significance of the results and contrast them with the relevant literature. The appendices contain the technical details.

\section{The tight-binding model \label{sec:tbmodel}}

In this section, we present the normal state real-space tight-binding model and the basic properties of monolayer TMDs. 

\subsection{Crystal and orbital structure} 

A 1H-TMD monolayer consists of a transition metal layer sandwiched between two chalcogen layers. Both the transition metal and chalcogen layers are triangular lattices intertwined with respect to each other. The view along the $c$-axis shows the resultant hexagonal lattice structure in prismatic coordination, see figure \ref{fig:crystal}a.  Hereafter we refer to the 1H-MX$_2$ structure as a TMD monolayer. The point-group symmetry of a TMD monolayer is $D_{3h}$, the symmetry of a triangle endowed with a basal mirror plane $\sigma_h$.
Although an inversion center lacks, several mirror planes exist; see figure \ref{fig:crystal}.
This changes in bilayer 2H-TMD's, where inversion is restored and the point group symmetry enlarges to $D_{3d}$.  

The Bloch states of TMD monolayers at the Fermi level, receive a dominant orbital contribution from the $4d$ transition metal orbitals, and the chalcogen $p$ orbitals contribute less \cite{Roldan2014,Liu2015,Silva-Guillen2016}. 
It is, therefore, possible to construct a low-energy three-orbital tight-binding model of TMD monolayers,
taking into account hopping only between transition metal $4d$ orbitals \cite{Mattheiss1973}. Within $D_{3h}$, one can use
the orbitals $d_{z^2}$, $d_{xy}$ and $d_{x^2-y^2}$ as a minimal basis set. Liu \textit{et al} constructed such a tight-binding model in momentum space \cite{Liu2013}. In this paper, we reformulate the momentum space model in real-space. The details are explained in appendix \ref{app:tb}. 
In our figures, we adopt the RGB color scheme $d_{z^2}$ (red), $d_{xy}$ (green) and $d_{x^2-y^2}$ (blue) to refer to the orbitals.

\subsection{The normal state Hamiltonian}

The normal state Hamiltonian $\mathcal{H}_\mathrm{N}$ contains the four terms
\begin{equation}
\mathcal{H}_\mathrm{N}=\mathcal{H}_0+\mathcal{H}_{\mathrm{SO}}+\mathcal{H}_\mathrm{Z}+\mathcal{H}_\mathrm{D},
\label{eq:model}
\end{equation}
where $\mathcal{H}_0$ is the bare tight-binding Hamiltonian, $\mathcal{H}_{\mathrm{SO}}$ contains the SOC interaction, $\mathcal{H}_\mathrm{Z}$ is the Zeeman term arising due to an in-plane magnetic field, and $\mathcal{H}_\mathrm{D}$ adds random on-site disorder. We detail each term below.

The bare tight-binding term is given by
\begin{equation}
\mathcal{H}_0=\sum_{\langle\mathbf{i},\mathbf{j}\rangle}\sum_{\mu,\nu,\sigma} t_{\mathbf{ij}}^{\mu\nu}c^\dag_{\mathbf{i}\mu\sigma}c_{\mathbf{j}\nu\sigma}+\sum_{\mathbf{i},\mu,\sigma}\left(\epsilon^\mu-\mu_0\right)c^\dag_{\mathbf{i}\mu\sigma}c_{\mathbf{i}\mu\sigma}.
\label{eq:h0}
\end{equation}
The operator $c^\dag_{\mathbf{i}\mu\sigma}$ creates an electron in orbital $\mu$ at an atomic transition metal site $\mathbf{i}$ with spin projection $\sigma$. The hopping amplitudes $t_{\mathbf{ij}}^{\mu\nu}$ are included up to third nearest-neighbors, and their values fit the band structure and orbital weights as calculated by first principle methods \cite{Liu2013,He2018}. 
We explain the details for obtaining all hopping amplitudes in real-space in appendix \ref{app:tb}. In the second term, $\epsilon^\mu$ is an on-site energy of the orbital $\mu$, and $\mu_0$ is the chemical potential.  

\begin{figure*}
\centering
\includegraphics[width=\textwidth]{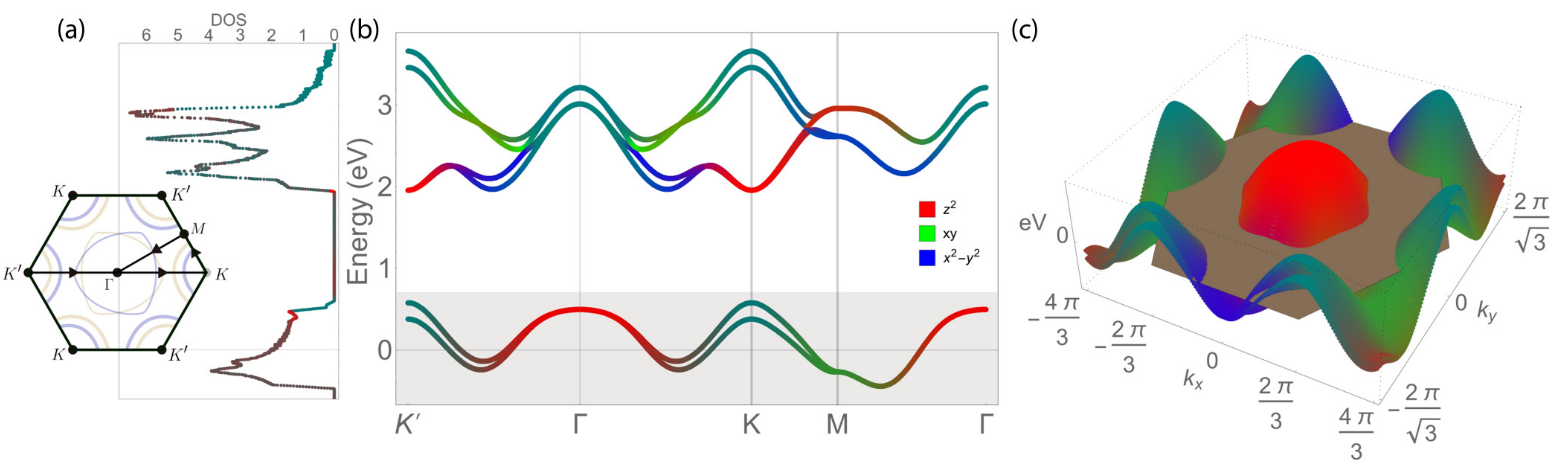}
\caption{\label{fig:bands} 
The electronic structure of the three-orbital tight-binding model for 1H-NbSe$_2$. 
a) The density of states of the band structure in (b). The first Brillouin zone inset shows a schematic projection of the Fermi surface. Gold bands have spin-projection $\uparrow$, and blue spin $\downarrow$. The arrows show the cuts plotted in (b). The spin-projection shows that while inversion is broken, time-reversal is respected. 
b) Band structure along high symmetry lines. The colors show the orbital content of the bands. SOC vanishes on the high symmetry line $\Gamma M$. 
c) 3D version of the band structure, focusing on the Fermi level crossing bands, which corresponds to the shaded grey region in (b). The brown hexagon delimits the first Brillouin zone at the Fermi level. }
\end{figure*}

The direction of the SOC magnetic induction $\mathbf{B}_\mathrm{SO}$ follows from the specific form of the electric crystal field $\nabla V$. 
Because the coordination of the chalcogen atoms respects the mirror $\sigma_h$, $\nabla V$ is confined to the in-plane direction.
Therefore, $\mathbf{B}_\mathrm{SO}\parallel \mathbf{p}\times \nabla V$ is anti-symmetric throughout the unit cell, where $\mathbf{p}$ is momentum. Thus, $\mathbf{B}_\mathrm{SO}$ is dubbed an \textit{Ising} SOC field, because it locks the spins in the out-of-plane direction, making them robust against in-plane magnetic fields \cite{Ugeda2015,Zhou2016,Bawden2016,DelaBarrera2018,Saito2016,Navarro-Moratalla2016,Xi2015}.  

In the tight-binding model, we include atomic SOC stemming from the heavy transition metal. 
The $z$-component of the orbital angular momentum operator $L_z$ acts on its eigenkets as $L_z|l,m\rangle=m|l,m\rangle$ ($\hbar=1$).
The $L_z$-eigenkets $|l,m\rangle$ are related to the orbital states $\{|d_{z^2}\rangle,|d_{xy}\rangle,|d_{x^2-y^2}\rangle\}$ by
\begin{equation}\label{eq:angular}
|d_{z^2}\rangle = |2,0\rangle;\quad  |2,\pm 2\rangle = \frac{1}{\sqrt{2}}\left(|d_{x^2-y^2}\rangle\pm i|d_{xy}\rangle\right).
\end{equation}
In the basis of $\{|d_{z^2}\rangle,|d_{xy}\rangle,|d_{x^2-y^2}\rangle\}$, the matrix representation of $L_z$ is
\begin{equation}
L_z = \begin{bmatrix}
0 & 0 & 0\\ 
0 & 0 & 2i\\ 
0 & -2i & 0
\end{bmatrix}.
\label{eq:half}
\end{equation} 
One can verify that all the elements of $L_x$ and $L_y$ are zero \cite{Roldan2014}.
Therefore, the SOC Hamiltonian involves the two in-plane orbitals \cite{Zhu2011} and can be written as
\begin{equation}
\mathcal{H}_\mathrm{SO} = i\lambda_\mathrm{SO}\sum_{\mathbf{i},\sigma,\sigma^\prime}\left(\sigma_z\right)_{\sigma\sigma^\prime} c^\dag_{\mathbf{i},x^2-y^2,\sigma}c_{\mathbf{i},xy,\sigma^\prime}+\mathrm{h.c.},
\label{eq:soc}
\end{equation}
where $\sigma_z$ is the third Pauli matrix.
Although this term is local, together with the parity lacking $\mathcal{H}_0$ provides an anti-symmetric splitting of the electronic states, while still preserving time-reversal symmetry. 
For the present form of the SOC, the $z$-component of the spin is a good quantum number. 
The specific form of \eqref{eq:soc} determines the structure of the induced triplet Cooper pair correlations.

The Zeeman term is responsible for the paramagnetic limiting effect and reads 
\begin{equation}
\mathcal{H}_\mathrm{Z}=-\frac{g\mu_\mathrm{B}}{2}\sum_{\mathbf{i},\mu}\sum_{\sigma,\sigma^\prime}\mathbf{B}\cdot\boldsymbol{\sigma}_{\sigma\sigma^\prime}c^\dag_{\mathbf{i}\mu\sigma}c_{\mathbf{i}\mu\sigma^\prime},
\end{equation}
where $\mu_\mathrm{B}$ is the Bohr magneton and $\boldsymbol{\sigma}=(\sigma_x,\sigma_y,\sigma_z)$ is the vector of Pauli matrices. For perpendicular magnetic fields applied to TMD monolayers, the $g$-factor is known to differ from its free electron value \cite{Kormanyos2014}. However, in this paper we examine only in-plane magnetic fields. Since $\mathbf{B}_\mathrm{SO}$ has no in-plane component, we adopt $g=2$ for simplicity \cite{Bauer2012}.

\subsubsection{On-site scalar disorder}

In this paper, we investigate the effect of the Anderson and dilute disorder.
In both cases, the disorder is realized as an on-site random scalar potential diagonal in the orbital index.
Such a short-range disorder is our way to model the intra-orbital elastic scattering with arbitrary scattering momenta. The on-site disorder Hamiltonian reads \cite{Lattices1956}
\begin{equation}
\mathcal{H}_\mathrm{D}= \sum_{\mathbf{i},\mu,\sigma} W_{\mathbf{i}}^\mu  c^\dag_{\mathbf{i}\mu\sigma}c_{\mathbf{i}\mu\sigma},
\label{eq:disorder}
\end{equation}
where $\{W_{\mathbf{i}}^{z^2},W_{\mathbf{i}}^{xy},W_{\mathbf{i}}^{x^2-y^2}\}=\{W_{\mathbf{i}}^{z^2},W_\mathbf{i},W_\mathbf{i}\}$, and both $W_{\mathbf{i}}^{z^2}$ and $W_\mathbf{i}$ are random disorder potentials. For Anderson disorder, the random potentials follow a uniform distribution in the interval $[-W/2,W/2]$. For dilute disorder, we use a Gaussian distribution for the disorder potentials with standard deviation $W$. In both cases, $W$ can be interpreted as the disorder strength. For our results, the choice of the probability distribution is immaterial, and we study Gaussian dilute disorder to contrast our results with the literature \cite{Ilic2017}.

For each realization of the on-site disorder $W_{\mathbf{i}}^{z^2}\propto W_\mathbf{i}$. Here we considered different realizations of the disorder  acting on $|d_{z^2}\rangle$ and on $\{|d_{x^2-y^2}\rangle,|d_{xy}\rangle\}$ orbitals as if they were independent.
We have verified that using a single realization for all the orbitals leads to the same results.
This can be explained as follows. 
The disorder is diagonal in orbital index. Moreover, the states crossing the Fermi level have either $|d_{z^2}\rangle$ or $\{|d_{x^2-y^2}\rangle,|d_{xy}\rangle\}$ orbital content.
Therefore, the two components of the disorder do not interfere.

In the dilute disorder scenario, disorder potential is present at a small fraction of randomly chosen sites at an impurity concentration $C_\mathrm{imp}=N_\mathrm{imp}/N_\mathrm{sites}\ll 1$. The scattering-rate of dilute disorder is then \cite{Wolfle2010,Ilic2017}
\begin{equation}
    \frac{\hbar}{\tau}=\pi\, C_\mathrm{imp}\,\rho(E_\mathrm{F})\left\langle W^2_\mathbf{i}\right\rangle,
    \label{eq:scatteringrate}
\end{equation}
where $\rho(E_\mathrm{F})$ is the density of states per unit cell per spin species at the Fermi level.

\section{Ising superconductivity \label{sec:unconventional}}

In this section, we comment on the specificities of unconventional Ising superconductivity, provide a group-theoretical analysis of on-site pairing correlations and present the superconducting interaction Hamiltonian.

In a single orbital system with both time-reversal and inversion symmetry, we can classify the superconducting phases by parity: either even-parity spin-singlet or odd-parity spin-triplet \cite{Sigrist1991}. In non-centrosymmetric systems such as monolayer TMD's, a definite parity lacks, and the superconducting states are parity-mixed \cite{Gorkov2001,Yip2014,Bauer2012,Smidman2017}. We denote pairing operators as $\hat{\Psi}$, expectation values as $\Psi\equiv \langle\hat{\Psi}\rangle$ and matrices whose elements are expectation values as $[\Psi]$. We refer to the $\Psi$'s as \textit{pairing correlations}, and to $\Delta=U\Psi$ as \textit{superconducting order parameters}, where $U$ is a pairing potential with dimension of energy, and $\Psi$ is dimensionless. 
We introduce a combined label $I=\{_{\mathbf{ij}}^{\mu\nu}\}$ for the lattice sites $\mathbf{i}$ and $\mathbf{j}$, and orbitals $\mu$ and $\nu$, such that a general pairing correlation $\Psi_{\mathbf{ij},\sigma\sigma^\prime}^{\mu\nu}=\langle c_{\mathbf{i}\mu\sigma}c_{\mathbf{j}\nu\sigma^\prime}\rangle$ is abbreviated as  $\Psi_{I,\sigma\sigma^\prime}$. We might use different bases for $[\Psi]$ matrices. We use the notation $[\Psi_{\mathbf{ij},\sigma\sigma^\prime}]$ for the orbital basis, and $[\Psi_I]$ for the spin basis.
Then, in the most general case, a superconducting pairing correlation in spin-space can be parametrized as 
\begin{align}
    [\Psi_I] & =\left(\psi_I+\mathbf{d}_I\cdot\boldsymbol{\sigma}\right)i\sigma_y=
\begin{bmatrix}
-d_{I,x}+id_{I,y} & \psi_I+d_{I,z} \\ 
-\psi_I+d_{I,z} & d_{I,x}+id_{I,y}
\end{bmatrix} 
\nonumber \\
    & =  
\begin{bmatrix}
\Psi_{I,\uparrow\uparrow} & \Psi_{I,\uparrow\downarrow} \\ 
\Psi_{I,\downarrow\uparrow} & \Psi_{I,\downarrow\downarrow} 
\end{bmatrix}. 
\label{eq:gap_matrix}
\end{align}
The matrix $[\Psi_I]$ can describe on-site ($\mathbf{i}=\mathbf{j}$) pairing, or Cooper pairs with the participating electrons placed at different sites $\mathbf{i}\neq\mathbf{j}$. 
In this paper, we compute on-site and nearest-neighbor pairing correlations. 
Similarly, equation \eqref{eq:gap_matrix} describes intra- and inter-orbital pairing for $\mu=\nu$ and $\mu\neq\nu$ respectively.
If either $\mathbf{d}_I=0$ or $\psi_I = 0$,  
the matrix $[\Psi_I]$ is anti-symmetric or symmetric in spin indices respectively.
In the single orbital case, the coexistence of both implies a lack of a definite parity. No such restriction exists in multi-orbital systems \cite{Bzdusek2017}.
In these systems, the SOC generically induces the triplets in addition to the singlets.

The singlet part of $\Psi_{I,\uparrow\downarrow}$ is odd under spin permutation $\psi_I\rightarrow -\psi_I$, and therefore the complex scalar function $\psi_I$ parametrizes singlet Cooper pairs that are even under the combined interchange of site and orbital indices, $(\mathbf{i}\mu) \leftrightarrow (\mathbf{j}\nu)$. 
The triplet order parameter is parametrized by a complex $d$-vector $\mathbf{d}_I=(d_{I,x},d_{I,y},d_{I,z})$. 
The $d$-vector is even in spin indices, and odd under the combined exchange of site and orbital indices, $(\mathbf{i}\mu) \leftrightarrow (\mathbf{j}\nu)$.
This implies that in the single orbital systems, the on-site superconducting correlations are necessarily singlet.
In the multi-orbital systems, the triplet correlations are allowed with the on-site order parameter which is odd under exchange of orbitals.

Whereas singlet Cooper pairs are strongly paramagnetically limited, triplets might have protected components against the action of a Zeeman field $\mathbf{B}$.
For strong SOC fields, the $d$-vector is parallel to the SOC magnetic induction $\mathbf{B}_\mathrm{SO}$ \cite{Frigeri2004,Ramires2016,Ramires2018,Smidman2017,Fischer2018}. 
Therefore, in the absence of external magnetic fields, the $d$-vector in monolayer TMDs is perpendicular to the layer with $\mathbf{d}_I=(0,0,d_{I,z})$. 
2D Ising superconductors have a superconducting triplet component in $\Psi_{I,\uparrow\downarrow}$ that is not suppressed by paramagnetic limiting. In the absence of an external magnetic field $\Psi_{I,\uparrow\uparrow}=\Psi_{I,\downarrow\downarrow}=0$. The application of a magnetic field or the presence of a substrate generating Rashba SOC will populate these terms. We show a schematic example in figure \ref{fig:tilt}.

The singlet and triplet channels are not decoupled from one another. Generally, quasi-particle excitation energies will depend on cross terms such as $\psi d_z^*+\psi^* d_z$, see appendix \ref{app:nodes} for an explicit example. This means that although $d_z$ is insensitive to in-plane magnetic fields, it is indirectly suppressed through the coupling with the singlet component $\psi$. Conversely, the presence of $d_z$ greatly enhances the paramagnetic limit $B_\mathrm{P}$. An enhanced critical field is reported in many experiments \cite{Xi2015,Bawden2016,Xing2017,DelaBarrera2018}.  

\subsection{
Group theoretic analysis of the on-site pairing correlations with and without SOC}
\label{sec:on-site}

Since our calculations are performed in real-space, it is instructive to derive the most generic form of the on-site correlations from the symmetry considerations.
We first discuss the case without SOC in section \ref{sec:on-site1}. 
In section \ref{sec:on-site2} we perform the analysis of the local Cooper correlations in the presence of SOC.
In both cases, our main focus is on the superconducting state that has the symmetry of the lattice, $A_1'$. 
This symmetric state is referred to as $s$-wave superconductivity for shortness.

\subsubsection{Local Cooper correlations in the absence of SOC}
\label{sec:on-site1}

In the absence of SOC, the wave function of the pair is a direct product of the orbital and spin wave functions.
As the total spin of a Cooper pair is a good quantum number, the spin part is either singlet or triplet.
It is, therefore sufficient to classify the orbital part of the on-site wave functions which has to be even in the case of spin singlet and odd in the case of the spin triplet.
The orbital part is classified in accordance with the $D_{3h}$ symmetry group.
The $\{|d_{z^2}\rangle \}$ orbital transforms as $A_1'$ while the two orbitals $\{|d_{xy}\rangle,|d_{x^2-y^2}\rangle \}$ transform as $E'$.
Clearly, the $\{|d_{z^2}\rangle \}$ orbital gives rise to the spin singlet, 
\begin{align}\label{singlets_no_SOC1}
\hat{\Psi}_{\mathbf{i}\mathbf{i}s_1}^{A'_1} & = \sum_{\sigma,\sigma'} (i \sigma_y)_{\sigma \sigma'}  c_{\mathbf{i} z^2 \sigma}c_{\mathbf{i} z^2 \sigma'}.      
\end{align}
The symmetric (anti-symmetric) part of the direct product $E'\otimes E'$ gives the singlets (triplets).
Referring to the character table \ref{tab:ch_double_D3H}, among the singlets there is one consistent with the $s$-wave symmetry of the superconducting state,
\begin{align}\label{singlets_no_SOC2}
\hat{\Psi}_{\mathbf{i}\mathbf{i}s_2}^{A'_1} = \sum_{\sigma,\sigma'} (i \sigma_y)_{\sigma \sigma'}  [ c_{\mathbf{i} xy \sigma}c_{\mathbf{i}xy \sigma'}\! +\! c_{\mathbf{i}x^2-y^2 \sigma}c_{\mathbf{i} x^2 - y^2 \sigma'} ]. 
\end{align}
In addition, we obtain a pair of spin singlet on-site correlations transforming as $E'$ analogous to the $d$-wave order parameter,  
\begin{align}
\label{singlets_no_SOC3}
\hat{\Psi}_{\mathbf{i}\mathbf{i}s_3,1}^{E'} & = \sum_{\sigma,\sigma'} (i \sigma_y)_{\sigma \sigma'}  [ c_{\mathbf{i} xy \sigma}c_{\mathbf{i}x^2-y^2 \sigma'} 
+
c_{\mathbf{i} x^2-y^2 \sigma}c_{\mathbf{i}xy \sigma'}]
\notag \\
\hat{\Psi}_{\mathbf{i}\mathbf{i}s_3,2}^{E'} & = \sum_{\sigma,\sigma'} (i \sigma_y)_{\sigma \sigma'}  [ c_{\mathbf{i} xy \sigma}c_{\mathbf{i}xy \sigma'} 
-
c_{\mathbf{i} x^2-y^2 \sigma}c_{\mathbf{i}x^2-y^2 \sigma'}].
\end{align}

The triplets are necessarily $A_2'$ symmetric,
\begin{align}\label{triplet_no_SOC4}
\hat{\Psi}_{\mathbf{i}\mathbf{i}t}^{A_2'}  = \sum_{\sigma,\sigma'} (\sigma_x)_{\sigma \sigma'}  [ c_{\mathbf{i} xy \sigma}c_{\mathbf{i}x^2-y^2 \sigma'}
-
c_{\mathbf{i} x^2-y^2 \sigma}c_{\mathbf{i}xy \sigma'}].
\end{align}
In the $s$-wave superconductor without SOC, only the combinations \eqref{singlets_no_SOC1} and \eqref{singlets_no_SOC2}  may acquire a finite expectation value.  
The triplet correlations, equation \eqref{triplet_no_SOC4} are not allowed.
In the next section, we demonstrate that the triplets are present along with singlets, once the SOC is turned on.

\subsubsection{Local Cooper correlations in the presence of SOC}
\label{sec:on-site2}

At finite SOC, the on-site orbital states split into three doublets, $\{|2,+2\uparrow \rangle, |2,-2\downarrow \rangle \}$, $\{|2,0\uparrow \rangle, |2,0\downarrow \rangle \}$ and $\{|2,-2\uparrow \rangle, |2,+2\downarrow \rangle \}$ transforming as $\bar{E}_1$, $\bar{E}_2$ and $\bar{E}_3$ respectively, see Table \ref{tab:ch_double_D3H}. 
As follows from the Table \ref{tab:ch_double_D3H}, each of the three doublets gives rise to exactly one combination of $s$-wave symmetry denoted below as $\hat{\bar{\Psi}}_{\mathbf{i}\mathbf{i}1}$, $\hat{\bar{\Psi}}_{\mathbf{i}\mathbf{i}2}$ and $\hat{\bar{\Psi}}_{\mathbf{i}\mathbf{i}3}$, respectively.
The $d_{z^2}$ orbital is unaffected by SOC and gives rise to the local correlation identical to equation \eqref{singlets_no_SOC1},
\begin{align}\label{OP_SOC1}
\hat{\bar{\Psi}}_{\mathbf{i}\mathbf{i}2}  = \hat{\Psi}_{\mathbf{i}\mathbf{i}s_1}^{A'_1}   \, .
\end{align}
The $s$-wave correlations constructed out of the other two orbitals, $d_{xy}$ and $d_{x^2-y^2}$, read
\begin{align}\label{OP_SOC2}
\hat{\bar{\Psi}}_{\mathbf{i}\mathbf{i}1}& = c_{\mathbf{i} 2+2; \uparrow }c_{\mathbf{i}2-2;\downarrow } - c_{\mathbf{i} 2+2; \downarrow }c_{\mathbf{i}2-2;\uparrow }
\notag  \\
\hat{\bar{\Psi}}_{\mathbf{i}\mathbf{i}3} & =  c_{\mathbf{i} 2-2; \uparrow }c_{\mathbf{i}2+2;\downarrow }-
c_{\mathbf{i} 2-2; \downarrow }c_{\mathbf{i}2+2;\uparrow }  \, .
\end{align}
The combinations listed in equations \eqref{OP_SOC1} and \eqref{OP_SOC2} condense in the $s$-wave superconductor.
Among the three combinations only the $ \hat{\bar{\Psi}}_{\mathbf{i}\mathbf{i}2}$, equation \eqref{OP_SOC1} derived from the $d_{z^2}$ orbitals is a pure spin singlet. 

The correlations $\hat{\bar{\Psi}}_{\mathbf{i}\mathbf{i}1(3)}$ in equation \eqref{OP_SOC2} contain singlet and triplet components.
These singlet and triplet combinations can be explicitly written using equation \eqref{eq:angular} as
\begin{equation} 
\label{SOC_s_t}
\hat{\bar{\Psi}}_{\mathbf{i}\mathbf{i}1}   - \hat{\bar{\Psi}}_{\mathbf{i}\mathbf{i}3}= \hat{\Psi}_{\mathbf{i}\mathbf{i}s_2}^{A_1'},\quad \hat{\bar{\Psi}}_{\mathbf{i}\mathbf{i}1}  + \hat{\bar{\Psi}}_{\mathbf{i}\mathbf{i}3}  = 
i \hat{\Psi}_{\mathbf{i}\mathbf{i}t}^{A_2'},
\end{equation}
respectively.
The factor of $i$ in front of the triplet component in equation \eqref{SOC_s_t} is required by the time-reversal invariance.
The coexistence of the singlet and triplet on-site correlations induced by the SOC is indeed verified numerically, see figure \ref{fig:correlations}.
It also follows from equation \eqref{SOC_s_t} that the $d$-vector introduced in equation \eqref{eq:gap_matrix} points out-of-the plane.
This can be traced to the horizontal mirror symmetry and has been confirmed numerically as well, see figure \ref{fig:correlations}.  

The above group theoretical considerations apply as is to the $\Gamma$ point in the reciprocal space. 
This implies the double degeneracy of the bands at $\Gamma$, see figure \ref{fig:bands}(b).
Indeed, the states at $\Gamma$ realize the double group of $D'_{3h}$ that has only two-dimensional spinor irreducible representations (irreps).
As another important implication of the symmetry, we point out the double degeneracy along a high symmetry $\Gamma M$ lines, see figure \ref{fig:bands}(a).
In this case, the two mirrors crossing along $\Gamma M$ ensure the vanishing of the SOC induced splitting. As a result, the Cooper pairs of electrons with momenta along $\Gamma M$ are pure spin singlets. 
As the SOC vanishes on $\Gamma M$, the external magnetic field generates nodes along $\Gamma M$ once the Zeeman splitting exceeds the superconducting gap. 
This is discussed in more details in section \ref{sec:disc_nodal}.  

\begin{table}
\begin{tabular}{c|c|c|c|c|c|c|c|c|c}
 $D_{3h}'$      &  $E$  &  $Q$ & $ 2 \sigma_h $  & $2 C_3 $  & $2 C_3^2 $ 
 & $2 s_3 $ & $2 Q s_3  $  & $ 6 \sigma_v $
  & $6 U_2 $  \\
   \hline
$A'_1$ & 1    &     1      &   1   &   1  &  1  &  1 & 1 & 1 & 1  \\
\hline
$A'_2$ &  1    &     1      &   1   &  1 & 1 & 1 & 1 & -1& -1  \\ 
\hline
$E'$     &  2    &     2     &   1   &  -1 &  -1  & -1 & -1 & 0 & 0 \\
\hline
\hline
$A''_1$ & 1    &     1      &   -1   &  1 & 1 & -1 & -1 & -1  & 1 \\
\hline
$A''_2$ & 1    &     1      &   -1   &  1 & 1 & -1 & -1 &  1  & -1 \\ 
\hline
$E''$ & 2    &     1      &   -1   &  -1 &  -1  & 1  & 1  & 0   & 0 \\
\hline 
\hline
\hline
$\bar{E}_1$   & 2    &    -2      &  0   &  1  & -1 & $\sqrt{3}$   & $-\sqrt{3}$ & 0 & 0   \\
\hline
$\bar{E}_2$ & 2    &    -2    &  0   &  1  & -1 &  $-\sqrt{3}$ & $\sqrt{3}$ & 0 & 0   \\
\hline
$\bar{E}_3 $ & 2    &    -2      &   0   & -2 & 2 &  0                 &  0  & 0 & 0   \\
\hline
\end{tabular}
\caption{The character table of the irreps of the double group $D_{3h}'$. 
Classes are listed in the first line. 
The $A_{1,2}'$, $A_{1,2}''$ and $E'$, $E''$ are vector irreps used to describe the symmetry properties of the $D_{3h}$ symmetric systems without SOC. The irreps labelled by the single (double) prime are even (odd) under $\sigma_h$.
The spinor irreps $\bar{E}_1$, $\bar{E}_2$ and $\bar{E}_3$ are two-dimensional and provide the description of the $D_{3h}$ symmetric systems with SOC. For these irreps the $2 \pi$ rotation around any axis, $Q=-1$. For vector irreps $Q=1$.
\label{tab:ch_double_D3H}}   
\end{table}

\subsection{The pairing interaction \label{sec:interaction}}

We consider the on-site attraction in the three orbital model. Here we neglect the SOC induced renormalization of the interaction term of the Hamiltonian and construct the latter ignoring SOC.

The interactions respecting the symmetry of the crystal are $A_{1}'$ scalars. 
We limit the consideration to the local, on-site interactions.
They are constructed by forming bi-linear combinations of the on-site correlations listed in section \ref{sec:on-site1} and give
\begin{widetext}
\begin{align}
\label{H_S_inv}
  \mathcal{H}_\mathrm{S} =  -\frac{1}{2} \sum_{\mathbf{i}} \Big[ & U^{z^2} (\hat{\Psi}_{\mathbf{i}\mathbf{i}s_1}^{A_1'})^{\dag} \hat{\Psi}_{\mathbf{i}\mathbf{i}s_1}^{A_1'}
  +U (\hat{\Psi}_{\mathbf{i}\mathbf{i}s_2}^{A_1'})^{\dag} \hat{\Psi}_{\mathbf{i}\mathbf{i}s_2}^{A_1'}+\left(U' (\hat{\Psi}_{\mathbf{i}\mathbf{i}s_1}^{A_1'})^{\dag}
\hat{\Psi}_{\mathbf{i}\mathbf{i}s_2}^{A_1'}  + \mathrm{h.c.}\right)+ \sum_{k=1,2}U'' (\hat{\Psi}_{\mathbf{i}\mathbf{i}s_3,k}^{E'})^{\dag} \hat{\Psi}_{\mathbf{i}\mathbf{i}s_3,k}^{E'} \notag \\
&+V (\hat{\Psi}_{\mathbf{i}\mathbf{i}t}^{A_2'})^{\dag} \hat{\Psi}_{\mathbf{i}\mathbf{i}t}^{A_2'} \Big].
\end{align}
\end{widetext} 
See appendix \ref{app:invariants} for proof. Equation \eqref{H_S_inv} is the most general form of the local pairing Hamiltonian allowed by the symmetry in the three orbital model.
It is rich enough to contain three $A_1'$ singlets $\{U^{z^2},U,U'\}$, one singlet of $E'$ symmetry ($U''$) and one triplet channel of $A_2'$ ($V$) symmetry. Since our goal is to study the emergence of the triplet correlations as an intrinsic property of the system, we set $V=0$ unless stated otherwise.
In addition, we study the $s$-wave superconductivity, and therefore set the attractive amplitude in the $E'$ channels to zero $U''=0$. This is legitimate as within the mean field approach channels of different symmetry decouple.
The coupling between the two $s$-wave singlets does not affect any of our results, and we set $U'=0$.

As a result, using \eqref{singlets_no_SOC1} and \eqref{singlets_no_SOC2} we rewrite \eqref{H_S_inv} as 

\begin{equation}
\begin{split}
\mathcal{H}_\mathrm{S}= & -U^{z^2}\sum_{\mathbf{i},\sigma}c^\dag_{\mathbf{i}z^2\sigma}c^\dag_{\mathbf{i}z^2,-\sigma}c_{\mathbf{i}z^2,-\sigma}c_{\mathbf{i}z^2,\sigma}\\
&-U\sum_{\mathbf{i},\sigma}\sum_{\alpha,\beta}c^\dag_{\mathbf{i}\alpha\sigma}c^\dag_{\mathbf{i}\alpha,-\sigma}c_{\mathbf{i}\beta,-\sigma}c_{\mathbf{i}\beta,\sigma},
\end{split}
\label{eq:interaction}
\end{equation}
where the orbital indices $\{\alpha,\beta\}$ run over the in-plane orbitals $\{d_{xy},d_{x^2-y^2}\}$. After mean-field decoupling we obtain
\begin{equation}
\begin{split}
\mathcal{H}_\mathrm{S} & =
\sum_{\mathbf{i}\sigma,\sigma^\prime} (i\sigma_y)_{\sigma\sigma^\prime}
\Delta_{\mathbf{i}\sigma\sigma^\prime}^{z^2}c^\dag_{\mathbf{i}z^2\sigma}c^\dag_{\mathbf{i}z^2\sigma^\prime}+\mathrm{h.c.}\\
& +\sum_{\mathbf{i}\sigma,\sigma^\prime}\sum_{\alpha,\beta} 
(i\sigma_y)_{\sigma\sigma^\prime}
\Delta_{\mathbf{i}\sigma\sigma^\prime}^{\alpha}c^\dag_{\mathbf{i}\beta\sigma}c^\dag_{\mathbf{i}\beta\sigma^\prime}+\mathrm{h.c.},
\end{split}
\label{eq:singlet_interaction}
\end{equation}
with the superconducting order parameters given by
\begin{equation}
\Delta_{\mathbf{i}\sigma\sigma^\prime}^\mu=(i\sigma_y)_{\sigma\sigma^\prime} U^\mu\langle c_{\mathbf{i}\mu\sigma}c_{\mathbf{i}\mu\sigma^\prime}\rangle = (i\sigma_y)_{\sigma\sigma^\prime} U^\mu\, \Psi_{\mathbf{ii},\sigma\sigma^\prime}^{\mu\mu},
\label{eq:gap_function}
\end{equation}
where $\mu$ runs over all orbitals and from \eqref{H_S_inv} we have $(U^{{z^2}},U^{{xy}},U^{{x^2-y^2}})=(U^{z^2},U,U)$. The superconducting order parameters \eqref{eq:gap_function} can be parametrized like \eqref{eq:gap_matrix}, and must be self-consistently determined. 
Although the pairing Hamiltonian \eqref{eq:singlet_interaction} only involves on-site intra orbital pair correlations $\Psi_{\mathbf{ii},\sigma\sigma^\prime}^{\mu\mu}$, the spin-locked normal state and more specifically SOC induces more general pairing correlations $\Psi_{\mathbf{ij},\sigma\sigma^\prime}^{\mu\nu}$ including inter-orbital pairs.

\section{Numerical method \label{sec:numerical}}

The model presented in the previous section is a real-space mean-field Bogoliubov-deGennes (BdG) Hamiltonian $\mathcal{H}_\mathrm{BdG}$. Since mean-field Hamiltonians are bi-linear in the operators, one can in principle find the eigenvalues and eigenvectors by exact numerical diagonalization. 
However, solving a matrix lattice Hamiltonian of large dimension $D$ using exact diagonalization methods rapidly turns into an intractable task. The dimension of the Hamiltonian matrix $\mathcal{H}_\mathrm{BdG}$ is determined by the degrees of freedom:
\begin{equation}
D=\mathrm{sites}\times\mathrm{orbitals}\times\mathrm{spins}\times\mathrm{electron/hole}.
\label{eq:dims}
\end{equation}
As an example, within an exact BdG approach with
$40\times 40$ atoms, one orbital, no SOC and electron-hole symmetry, the corresponding matrix Hamiltonian has dimension $D=(40\times 40)\times 1\times 1\times 2=3200$.  
Such matrices are still tractable for self-consistent exact diagonalization. In reference \cite{Gastiasoro2016}, exact-diagonalization is performed on a matrix with dimension $D=9000$.
In our multi-orbital system with SOC, a real-space $40\times 40$ lattice has a Hamiltonian of dimension $D=(40\times 40)\times 3\times 2\times 2=19200$. 
Doing self-consistent calculations on $19200\times 19200$ matrices using exact diagonalization is very time consuming and memory expensive on typical Desktop computers. 
Solving such matrix sizes (or even much bigger) is feasible within a Chebyshev Green's function expansion approach, even on a Desktop computer. In the following sections we summarize the main elements of the method, and for a detailed account, we refer the reader to the references \cite{Weisse2005b,Covaci2010,Nagai2011,Berthod2016}.

\subsection{Chebyshev expansion of the Green's function}

We are interested in the retarded Green's function $G(E)=(E-\mathcal{H}_\mathrm{BdG})^{-1}$ evaluated immediately above the real axis. A common choice of expansion polynomials are the Chebyshev polynomials of the first kind defined by $T_n(x)=\cos (n\arccos x)$, where $x\in[-1,1]$.  They are known for their good convergence properties and recursive relation $T_{n+1}=2x T_n(x)-T_{n-1}(x)$, with $T_0(x)=1$ and $T_1(x)=x$.
Because the Chebyshev polynomials are defined on $[-1,1]$, one has to rescale $\mathcal{H}_\mathrm{BdG}$ into the dimensionless form $\tilde{\mathcal{H}}_\mathrm{BdG}=(\mathcal{H}_\mathrm{BdG}-\mathfrak{b}\hat{\mathrm{1}})/\mathfrak{a}$, where $\mathfrak{a}$ is an upper bound estimate for the energy spectrum, and $\mathfrak{b}$ is the center of the spectrum.
We indicate all rescaled quantities with a tilde. Similarly, $\tilde{E}=(E-\mathfrak{b})/\mathfrak{a}$. Since BdG Hamiltonians have built-in electron-hole symmetry, $\mathfrak{b}=0$ and $\mathfrak{a}\approx E_\mathrm{max}$.  One can expand the retarded Green's function in terms of the $T_n(\tilde{\mathcal{H}}_\mathrm{BdG})$ as \cite{Weisse2005b,Covaci2010,Nagai2011,Berthod2016}
\begin{equation}
\begin{split}
&G(E+i0)=\frac{1}{E-\mathcal{H}_\mathrm{BdG}+i0}= \\
&-\frac{1}{\mathfrak{a}}\frac{i}{\sqrt{1-\tilde{E}^2}}\sum_{n=0}^\infty(2-\delta_{n0})T_n(\tilde{\mathcal{H}}_\mathrm{BdG})e^{-in\arccos\tilde{E}}.
\end{split}
\label{eq:expansion}
\end{equation}
In practice, one truncates the series at an expansion order $N-1$ and the Green's function $G(E+i0)$ and the Hamiltonian $\mathcal{H}_\mathrm{BdG}$ are projected onto a basis involving sites, orbitals and spins, such that the argument of the $T_n(x)$'s are $D$-dimensional matrices. The truncation introduces spurious oscillations in the Green's function known as Gibbs oscillations \cite{Weisse2005b}. To correct for these oscillations, guarantee the positivity of the poles, and improve convergence, we include the Jackson kernel in the summand of equation \eqref{eq:expansion} defined as
\begin{equation}
g_n=\frac{(N-n+1)\cos\left(\frac{n\pi}{N+1}\right)+\sin\left(\frac{n\pi}{N+1}\right)\cot\left(\frac{\pi}{N+1}\right)}{N+1}.
\label{eq:jackson}
\end{equation}
Other kernels with different convergence properties also can be used \cite{Weisse2005b,Berthod2016}. 

\subsection{Recursive implementation and resolution \label{sec:resolution}}

We wish to determine matrix elements of the retarded Green's function $\langle\alpha|G(E+i0)|\beta\rangle$. 
This amounts to the evaluation of the expansion moments $\langle\alpha|T_n(\tilde{\mathcal{H}}_\mathrm{BdG})|\beta\rangle=\langle\alpha|\psi_n\rangle$. With the starting vectors $|\psi_0\rangle=T_0(\tilde{\mathcal{H}}_\mathrm{BdG})=|\beta\rangle$ and $|\psi_1\rangle=T_1(\tilde{\mathcal{H}}_\mathrm{BdG})=\tilde{\mathcal{H}}_\mathrm{BdG}|\beta\rangle$, all vectors $|\psi_n\rangle$ up an arbitrary order $N-1$ can be recursively obtained using the relation $|\psi_{n+1}\rangle=2\tilde{\mathcal{H}}_\mathrm{BdG}|\psi_n\rangle-|\psi_{n-1}\rangle$. The core operation of the algorithm is then a sparse matrix vector multiplication of the type $\tilde{\mathcal{H}}_\mathrm{BdG}|\psi\rangle$, a process that can be efficiently parallelized. This shows the main advantage of the method as opposed to exact diagonalization.

The Chebyshev expansion, however, suffers from a drawback related to the resolution around the Fermi level $E=0$. Because the method requires one to rescale the entire spectral range of $\tilde{\mathcal{H}}_\mathrm{BdG}$, the resolution is set by $\mathfrak{a}$. Also, the zero's of the Chebyshev polynomials $T_n(\tilde{E})$ are sparser around $E=0$ (where high resolution is needed), and denser around the spectrum ends of $[-1,1]$. Therefore, if the energy scale of the superconducting gap centered at $E=0$ is $\Delta\ll\mathfrak{a}$, one has to guarantee that the expansion order $N$ is sufficiently large to resolve the smallest energy scale of interest: in this case $\Delta$. More precisely, the zeros of $T_n(\tilde{E})$ are $\tilde{E}_k=\cos[\pi/N(k+1/2)]$ such that the least resolution around $\tilde{E}=0$ is given by (for $N$ odd)
\begin{equation}
\frac{\Delta E}{\mathfrak{a}}=\tilde{E}_{(N-1)/2}-\tilde{E}_{(N-3)/2}=\sin\frac{\pi}{N}\approx \frac{\pi}{N}.
\label{eq:resolution}
\end{equation}
Therefore, the bandwidth-gap ratio $\mathfrak{a}/\Delta$ should be of the same order of $N$. We expand on more specific computational details in appendix \ref{app:kpm}.

\subsection{LDOS and superconducting gaps}

We define normal and anomalous (superconducting) components of the retarded Green's function respectively as
\begin{align}
G_{\mathbf{ij},\sigma\sigma^\prime}^{\mu\nu}(E+i0) & =\langle c_{\mathbf{i}\mu\sigma}(E-\mathcal{H}_\mathrm{BdG}+i0)^{-1}c_{\mathbf{j}\nu\sigma^\prime}^\dag\rangle;\\
F_{\mathbf{ij},\sigma\sigma^\prime}^{\mu\nu}(E+i0) & =\langle c_{\mathbf{i}\mu\sigma}(E-\mathcal{H}_\mathrm{BdG}+i0)^{-1}c_{\mathbf{j}\nu\sigma^\prime}\rangle,
\label{eq:normalG}
\end{align}  
and a similar anomalous component involving creation operators is omitted.
The electronic local density of states (LDOS) relates to \eqref{eq:normalG} via
\begin{equation}
\rho_\mathbf{i}(E)=-\frac{1}{\pi}\sum_{\mu,\sigma}\mathrm{Im}\,G_{\mathbf{ii},\sigma\sigma}^{\mu\mu}(E+i0).
\label{eq:ldos}
\end{equation}
From this we also can extract the partial orbital contributions $\rho_{\mathbf{i}\mu}(E)$ to the LDOS.  

A general pairing correlation $\Psi_{\mathbf{ij},\sigma\sigma^\prime}^{\mu\nu}=\langle c_{\mathbf{i}\mu\sigma}c_{\mathbf{j}\nu\sigma^\prime}\rangle$ is related to the anomalous Green's function via  \cite{Berthod2016}
\begin{align}
\label{eq:gap}
&\Psi_{\mathbf{ij},\sigma\sigma^\prime}^{\mu\nu}=\\
&i\int_{-\infty}^\infty \frac{\mathrm{d}E}{2\pi} f(E)\left[F_{\mathbf{ij},\sigma\sigma^\prime}^{\mu\nu}(E+i0)-F_{\mathbf{ij},\sigma\sigma^\prime}^{\mu\nu}(E-i0)\right], \nonumber 
\end{align} 
where $f(E)=(e^{E/k_B T}+1)^{-1}$ is the Fermi distribution. Unlike the LDOS that is always a real quantity, the $\Psi_{\mathbf{ij},\sigma\sigma^\prime}^{\mu\nu}$ involve the evaluation of the Green's function in both upper and lower complex half planes. However, due to electron-hole symmetry of the BdG Hamiltonian, one can show that the anomalous Green's function and the matrix elements of $T_n(\tilde{\mathcal{H}}_\mathrm{BdG})$ have the properties
\begin{align}
\label{eq:hole1}
F_{\mathbf{ij},\sigma\sigma^\prime}^{\mu\nu}(E-i0) & =F_{\mathbf{ji},\sigma^\prime\sigma}^{\nu\mu}(-E+i0);\\
\langle c_{\mathbf{i}\mu\sigma}|T_n(\tilde{\mathcal{H}}_\mathrm{BdG})|c_{\mathbf{j}\nu\sigma^\prime}\rangle & =(-1)^{n+1}\langle  c_{\mathbf{j}\nu\sigma^\prime}|T_n(\tilde{\mathcal{H}}_\mathrm{BdG})|c_{\mathbf{i}\mu\sigma}\rangle. \nonumber 
\end{align}
These properties are derived in appendix \ref{app:electronhole}.
With this, equation \eqref{eq:gap} simplifies to 
\begin{equation}
\label{eq:gapB}
\Psi_{\mathbf{ij},\sigma\sigma^\prime}^{\mu\nu}  =\sum_{n=1}^\infty D_n\mu_n,\quad \mu_n  =\langle c_{\mathbf{i}\mu\sigma}|T_n(\tilde{\mathcal{H}}_\mathrm{BdG})|c_{\mathbf{j}\nu\sigma^\prime}\rangle,
\end{equation} 
where the zeroth order expansion moment has dropped out from the sum because $\mu_0=0$\footnote{The zeroth order expansion moment should not be mistaken with the chemical potential.}, and
all the information about temperature is contained in
\begin{equation}
\begin{split}
D_n &=\frac{2}{\pi}\int_{-1}^{1}\mathrm{d}\tilde{E}\tilde{f}(\tilde{E})\frac{\cos\left( n\arccos\tilde{E}\right)}{\sqrt{1-\tilde{E}^2}}\\
&\approx \frac{2}{N}\sum_{k=0}^{N-1}\tilde{f}(\tilde{E}_k)\cos\frac{n\pi}{N}\left(k+\frac{1}{2}\right),
\end{split} 
\label{eq:dn}
\end{equation}
where $\tilde{E}_k=\cos[\pi/N(k+1/2)]$ are the Chebyshev abscissas and a Chebyshev-Gauss quadrature was used to obtain the second line (see appendix \ref{app:kpm}). At $T=0$ the integral or the sum in equation \eqref{eq:dn} can be evaluated analytically.
The coefficients $D_n$ are supplemented
with the Jackson kernel defined in equation \eqref{eq:jackson}. The second line of equation \eqref{eq:dn} has the form of a Cosine Fourier transform, such that one can use a fast Fourier transform algorithm to perform the integral efficiently.  

If $\tilde{\mathcal{H}}_\mathrm{BdG}$ is real, then the list of expansion moments $\{\mu_n\}$ is also real and no imaginary part of $\Psi_{\mathbf{ij},\sigma\sigma^\prime}^{\mu\nu}$ develops. However, if $\tilde{\mathcal{H}}_\mathrm{BdG}$ contains imaginary elements, such as coming from SOC, $\Psi_{\mathbf{ij},\sigma\sigma^\prime}^{\mu\nu}$ might develop an imaginary part accordingly. 
For a pairing interaction in the $A_1'$ channel in \eqref{H_S_inv}, only the elements $\Psi_{\mathbf{ii},\sigma\sigma^\prime}^{\mu\mu}$ are self-consistently updated and converged. Once the $\Psi_{\mathbf{ii},\sigma\sigma^\prime}^{\mu\mu}$ converged, one can probe any pairing correlation $\Psi_{\mathbf{ij},\sigma\sigma^\prime}^{\mu\nu}$.  

\begin{figure*}
\centering
\includegraphics[width=0.9\textwidth]{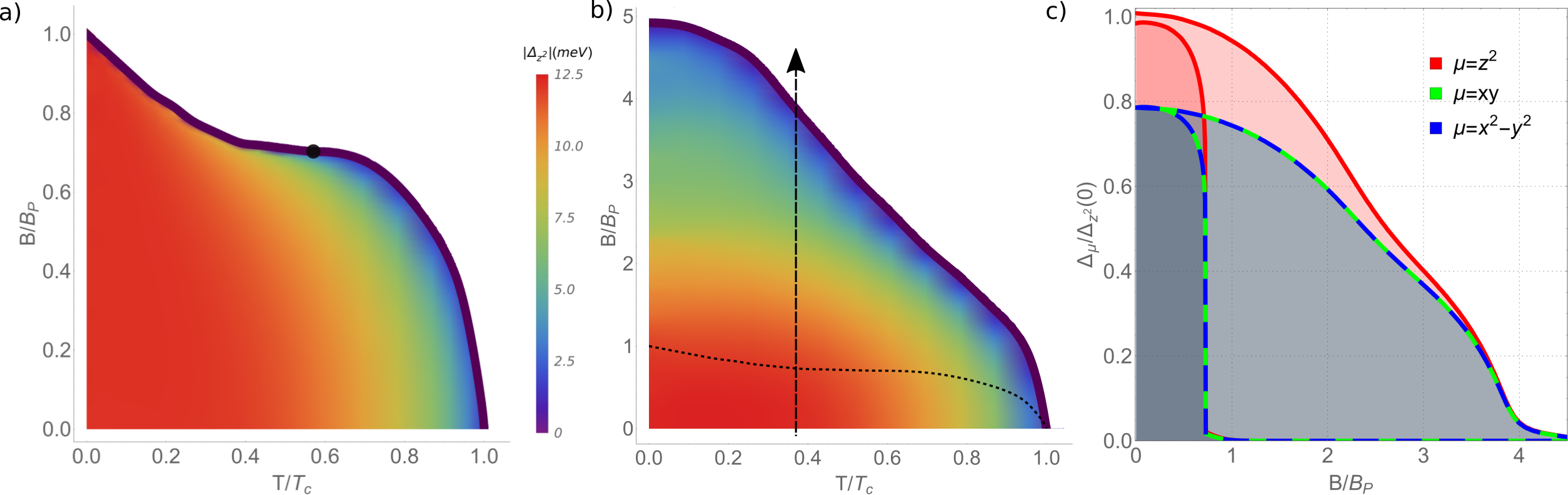}
\caption{\label{fig:diagrams} 
Magnetic induction-temperature superconducting phase diagrams showing the
enhancement of the paramagnetic critical field due to Ising SOC. 
a) The case for $\lambda_\mathrm{SO}=0$. Above the point at $T^*\approx 0.57 T_c$ the transition changes from first-order to second-order. The black point indicates the tricritical point. 
b) Phase diagram with $\lambda_\mathrm{SO}/\Delta_{z^2}(0)\approx 17$, showing strong enhancement of the critical magnetic field. Here $B_\mathrm{P}$ and $T_c$ are taken from the case with $\lambda_\mathrm{SO}=0$. The dashed line shows the paramagnetic transition line obtained in (a), and the arrow indicates the constant temperature cut that is analyzed throughout the paper.
c) Constant temperature cut of the phase diagrams at $T\approx 0.37 T_c$. The curves dropping below $B_\mathrm{P}$ correspond to the case without SOC, and the curves dropping at $B\approx4B_\mathrm{P}$ show the enhancement due to SOC. $\Delta_{z^2}(0)$ is the maximum value of the superconducting gap of the $4d_{z^2}$ orbital for the $\lambda_\mathrm{SO}=0$ case.
}
\end{figure*}

\section{Results \label{sec:results}}

In this section, we present the results of our numerical simulations. We show magnetic field-temperature phase diagrams with and without SOC demonstrating the enhancement of the critical field, find the on-site and nearest-neighbor superconducting pairing correlations, and explore the effects of on-site scalar impurities on the phase diagram. 

\subsection{Magnetic field--temperature phase diagrams}

\subsubsection{Setting up the parameters without SOC} 

We first examine the clean case ($W=0$) without SOC ($\lambda_\mathrm{SO}=0$) and use it as a reference system. Applying a magnetic field $B$ in the $x$ direction, the Hamiltonian $  \mathcal{H}_\mathrm{BdG}=\mathcal{H}_\mathrm{N}+\mathcal{H}_\mathrm{S}$ contains only real matrix elements.
It then follows from \eqref{eq:gapB} that no imaginary expansion moments $\mu_n$ can be generated, and as a consequence the pairing correlations $\Psi_{\mathbf{ij},\sigma\sigma^\prime}^{\mu\nu}$ are real.
Because singlet and triplet components transform as $\psi_I\rightarrow \psi_I^*$ and $\mathbf{d}_I\rightarrow -\mathbf{d}_I^*$ under time-reversal-operation,
no triplet correlations can be induced without SOC at zero magnetic fields. 

We assume an attractive interaction in the $A_1'$ channel, as introduced in \eqref{H_S_inv} with $U^{z^2}=U$. In principle, one can adjust $U$ such that the order parameters vanish at the system's superconducting critical temperature $T_c$. 
In monolayer 1H-NbSe$_2$, $T_c\approx 2$ K \cite{Ugeda2015,Khestanova2018}. This yields a BCS zero temperature gap $\Delta(0)= 1.76 k_\mathrm{B} T_c\approx 0.3$ meV, and the ratio between the SOC and superconducting energy scales is estimated to be of order $\lambda_\mathrm{SO}/\Delta(0)\sim 200$ \cite{DelaBarrera2018}.  
To be able to accurately calculate energy scales below $1$ meV, the resolution of the Chebyshev expansion method should have $\mu$eV precision. This means that according to \eqref{eq:resolution}, the number of Chebyshev expansion moments would have to be of order $N=\pi\mathfrak{a}/(\Delta E)\sim 10^7$. This is intractable if one needs to do a large amount of self-consistent calculations. 
Therefore, in this paper, we use $N=10^4$ (unless explicitly stated), which allows us to accurately resolve energy scales above $0.1$ meV and examine ratios up to $\lambda_\mathrm{SO}/\Delta(0)\approx 17$. 
Although the SOC-gap ratio is typically an order of magnitude larger in monolayer TMD's, we are still operating in the regime where $\Delta <\lambda_\mathrm{SO}\lesssim E_\mathrm{F}$, where $E_\mathrm{F}$ is the Fermi energy. Therefore, we use $U=0.42$ eV, which yields the largest zero temperature superconducting gap of $\Delta_{z^2}(0)\approx 12$ meV with $\lambda_\mathrm{SO}=0$. All subsequent plots involving superconducting order parameters are normalized with respect to $\Delta_{z^2}(0)$. 

To perform the numerical simulations, we consider a $40\times 40$ triangular lattice with periodic boundary conditions. 
Starting with random initial conditions,
the superconducting order parameters \eqref{eq:gap_function} converge self-consistently below $10$ $\mu$eV precision.  
In figure \ref{fig:diagrams}(a) we map out a magnetic induction-temperature phase diagram with $\lambda_\mathrm{SO}=W=0$. 
At $T=0$, we obtain a paramagnetic critical field of $B_\mathrm{P}\approx 250$ T. 
We use this value as normalization in subsequent plots. The color gradient shows the amplitude of $\Delta_{\mathbf{i}\uparrow\downarrow}^{z^2}=U\langle c_{\mathbf{i}z^2\uparrow}c_{\mathbf{i}z^2\downarrow}\rangle$ (abbreviated as $\Delta_{z^2}$). The color gradients for the other order parameters $\Delta_{xy(x^2-y^2)}$ are very similar, and for this reason, we only show $\Delta_{z^2}$. Below the temperature $T^*\approx 0.56 T_c$, the transition line is of first-order, and above $T^*$ it is of second-order \cite{Matsuda2007}. Inside the critical field transition line, the superconducting order parameters only vary appreciably above $T^*$.

\subsubsection{With SOC: enhancement of the critical magnetic field} 

We now examine the case where $\lambda_\mathrm{SO}$ is an order of magnitude larger than the superconducting energy scales, and for this, we set $\lambda_\mathrm{SO}=0.2$ eV, such that $\lambda_\mathrm{SO}/\Delta_{z^2}(0)\approx 17$. The phase diagram in figure \ref{fig:diagrams}(b) shows a five-fold enhancement of the paramagnetic critical magnetic field with respect to the $\lambda_\mathrm{SO}=0$ case.
This enhancement is due to the form of the Ising SOC interaction \eqref{eq:soc}, which locks the spins in the out-of-plane direction, and makes superconductivity robust against in-plane magnetic fields.
Many experiments report an enhancement of the upper critical field in Ising superconductors, and might even be more exaggerated with respect to our simulations depending on the TMD family, because of the larger $\lambda_\mathrm{SO}/\Delta$ ratio \cite{Xi2015,Navarro-Moratalla2016,Saito2016,DelaBarrera2018}.

Unlike the $\lambda_\mathrm{SO}=0$ case, the critical transition line is always of second-order \cite{Sohn2018}, in stark contrast to ordinary paramagnetically limited superconducting thin films \cite{Meservey1975,Meservey1994,Adams1998,Zocco2013}. From the finite temperature cross-section at $T=0.37 T_c$ in figure \ref{fig:diagrams}c, one can clearly identify the first-order phase transition without SOC, and the second-order phase transition with Ising SOC. Our numerical calculations show that $\Delta_{xy}=\Delta_{x^2-y^2}$ holds, which is a requirement of symmetry imposed by the $A_1'$ pairing interaction \eqref{eq:interaction}.

\subsection{Pairing correlations} 

Real-space BdG theory benefits from the ability to self-consistently find the superconducting pairing correlations $\Psi_{\mathbf{ij},\sigma\sigma^\prime}^{\mu\nu}=\langle c_{\mathbf{i}\mu\sigma}c_{\mathbf{j}\nu\sigma^\prime}\rangle$ that emerges from the spin-locked normal state. 
We will show that no other attractive pairing channel other than $A_1'$ is necessary to induce triplet correlations. We choose to probe for on-site and nearest neighbor (nn) superconducting correlations.

For each correlation, whether it is local or of nearest-neighbor type,
the matrix $[\Psi_I]$ in spin-space introduced in \eqref{eq:gap_matrix} has a spectral form
\begin{align}
[\Psi_I][\Psi_I]^\dag = & +\left(|\psi_I|^2+|\mathbf{d}_I|^2\right)\sigma_0\nonumber \\
&+\bigr(\underbrace{\psi_I\mathbf{d}_I^*+\psi_I^*\mathbf{d}_I}_{\xcancel{\mathcal{I}}}+i\underbrace{\mathbf{d}_I\times\mathbf{d}_I^*}_{\xcancel{\mathcal{T}}}\bigr)\cdot\boldsymbol{\sigma}.
\label{eq:nonunitary}
\end{align}
Here $\sigma_0$ is the $2\times 2$ unit matrix.
This shows the explicit coupling of the singlet and triplet components, and potentially a net Cooper pair spin polarization $i\mathbf{d}_I\times\mathbf{d}_I^*$. The first line \eqref{eq:nonunitary} proportional to $\sigma_0$ is unitary. The coupling $\psi_I\mathbf{d}_I^*+\psi_I^*\mathbf{d}_I$ is non-unitary in inversion $\mathcal{I}$ and $i\mathbf{d}_I\times\mathbf{d}_I^*$ in non-unitary in time-reversal $\mathcal{T}$ \cite{Yoshida2014}. 


Writing the time-reversal operator as $\mathcal{T}=i\sigma_y\mathcal{K}$, where $\mathcal{K}$ is the conjugation operator; by comparing $[\Psi_I]$ to $\mathcal{T}[\Psi_I]\mathcal{T}^{-1}$, we infer that under time-reversal operation $\psi_I\rightarrow \psi_I^*$ and $\mathbf{d}_I\rightarrow -\mathbf{d}_I^*$. 
With this, one can verify that $i\mathbf{d}_I\times\mathbf{d}_I^*$ is the only term that breaks time-reversal in \eqref{eq:nonunitary}. 
Therefore, for the discussion in real-space,
the superconducting state remains time-reversal-symmetric if $\psi_I$ is real, and the $d$-vector components are purely imaginary.   

Our simulations reveal that $\Psi_{I,\uparrow\downarrow}$ remains time-reversal-symmetric even in the presence of an in-plane magnetic field. A magnetic field induces real $d_{I,x}$ and $d_{I,y}$ in $\Psi_{I,\uparrow\uparrow}$ and $\Psi_{I,\downarrow\downarrow}$, whereas $d_{I,z}$ remains purely imaginary. In the following sections we discuss time-reversal-symmetric and time-reversal-breaking (TRB) pairing correlations separately.  
 
\begin{figure*}
\centering
\includegraphics[width=0.98\textwidth]{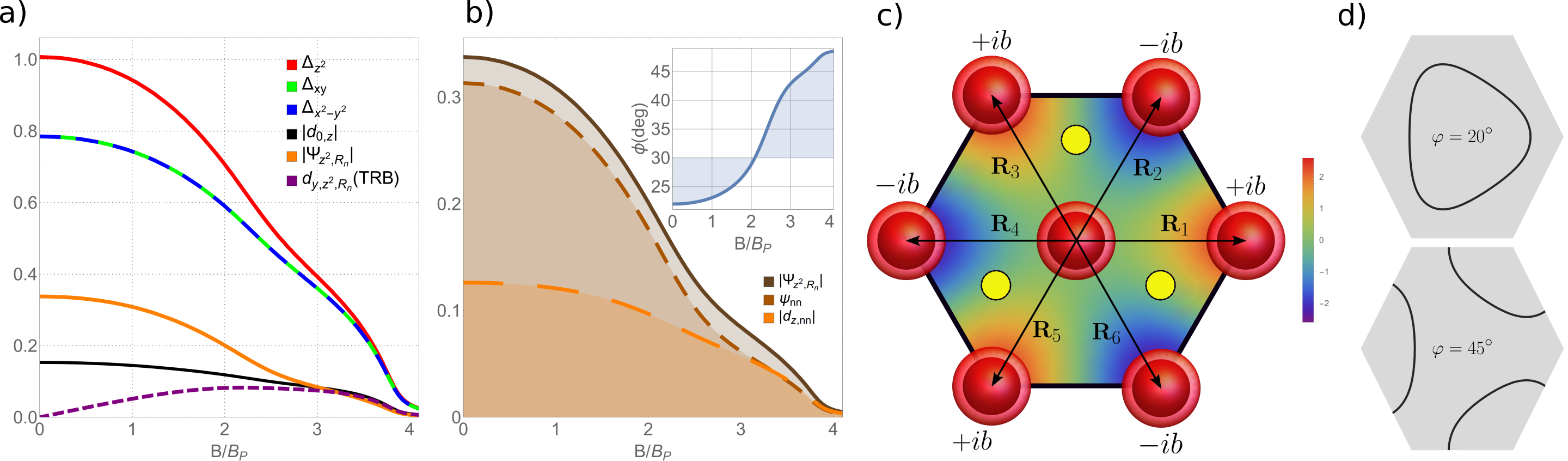}
\caption{\label{fig:correlations} 
Magnetic field dependence of the
superconducting pairing correlations at $T\approx 0.37T_c$.  
a) The three superconducting on-site order parameters $\Delta_{z^2}$ (red), $\Delta_{xy}$ (green) and $\Delta_{x^2-y^2}$ (blue). Among the many induced correlations, here we show the on-site inter-orbital triplet $d_{0,z}$ (black), the nearest-neighbor correlation $\Psi^{z^2}_{\mathbf{R}_n}$ along $\mathbf{R}_n$ (orange) for some fixed $n$, and the time-reversal-breaking (TRB) on-site component $d_{\mathbf{R}_n,y}^{z^2}$ (dashed-violet) induced by the $x$-directed magnetic field. The order parameters (indicated by $\Delta$) have dimension of energy, but the correlations are dimensionless.
b) The singlet $\mathrm{Re}\,\Psi^{z^2}_{\mathbf{R}_n}=\psi_\mathrm{nn}$ and triplet parts $|\mathrm{Im}\,\Psi^{z^2}_{\mathbf{R}_n}|=|d_{\mathrm{nn},z}|$. The evolution of $\varphi=\arctan(|d_{\mathrm{nn},z}|/\psi_\mathrm{nn})$ shows the increasing imbalance between the singlet and triplet components with increasing field.  
c) Nearest-neighbor modulation of the induced triplet component. The color bar shows the anti-symmetric sign modulation of the triplet component of $\Psi^{z^2}_\mathrm{nn}(\mathbf{k})$. 
d) The nodal line of $\Psi^{z^2}_\mathrm{nn}(\mathbf{k})$ in the first Brillouin zone for $\varphi=20^\circ$ and $\varphi=45^\circ$. The transition in topology of the nodal line occurs at $\varphi_c\approx 30^\circ$.  
}
\end{figure*}

\subsubsection{Time-reversal symmetric correlations}

From the discussion in section \ref{sec:unconventional}, we already know that at zero magnetic field correlations of the type $\Psi_{I,\uparrow\uparrow}=\Psi_{I,\downarrow\downarrow}=0$. This is numerically verified. We now look at the on-site sub-matrix of $\Psi_{I,\uparrow\downarrow}$ in the orbital basis $\{d_{z^2},d_{xy},d_{x^2-y^2}\}$, and write
\begin{align} 
\label{eq:onsite_block}
[\Psi_{\mathbf{ii},\uparrow\downarrow}] &=
\begin{bmatrix}
\Psi_{z^2} & \Psi_{xy,z^2} & \Psi_{x^2-y^2,z^2}\\
\Psi_{z^2,xy} & \Psi_{xy} & \Psi_{x^2-y^2,xy}\\
\Psi_{z^2,x^2-y^2} & \Psi_{xy,x^2-y^2} & \Psi_{x^2-y^2}
\end{bmatrix} \nonumber  \\
& = \begin{bmatrix}
\Psi_{z^2} & 0 & 0\\ 
0 & \Psi_{xy} & d_{0,z}\\ 
0 & -d_{0,z} & \Psi_{x^2-y^2}
\end{bmatrix}.
\end{align} 
The site and spin indices of the matrix elements were omitted and are implicit for clarity, that is for instance, $\Psi_{\mathbf{ii},\uparrow\downarrow}^{z^2}=\Psi_{z^ 2}$ and $d_{0,z}=d_{\mathbf{ii},z}^{x^2-y^2,xy}$.
The diagonal intra-orbital pairing correlations in \eqref{eq:onsite_block} are the only ones that form the superconducting order parameters $\Delta_{\mathbf{i}\sigma\sigma^\prime}^\mu=(i\sigma_y)_{\sigma\sigma^\prime} U^\mu \Psi_{\mathbf{ii},\sigma\sigma^\prime}^{\mu\mu}$, which converge self-consistently. After reaching self-consistent convergence for the (necessarily real and singlet) diagonal elements, we probe the content of the off-diagonal elements. We find that for finite SOC, $\Psi_{x^2-y^2,xy}=-\Psi_{xy,x^2-y^2}=d_{0,z}$ is purely imaginary, which corresponds to the $A_2'$ symmetric triplet state \eqref{triplet_no_SOC4}.
This is also known as an orbital-singlet state \cite{Fu2010,Michaeli2012}. Because the SOC matrix \eqref{eq:half} does not involve $d_{z^2}$, the matrix elements elements in \eqref{eq:onsite_block} involving $d_{z^2}$ are zero.
This induced triplet component clearly reflects the structure of the angular momentum matrix $L_z$ \eqref{eq:half}.

Similarly, again after converging the order parameters $\Delta_{\mathbf{i}\sigma\sigma^\prime}^\mu$, we probe the contents of the nearest-neighbor pairing correlations
\begin{equation} 
[\Psi_{{\mathbf{R}_n},\uparrow\downarrow}] =
\begin{bmatrix}
\Psi^{z^2}_{\mathbf{R}_n} & \Psi^{xy,z^2}_{\mathbf{R}_n} & \Psi^{x^2-y^2,z^2}_{\mathbf{R}_n}\\ 
\Psi^{z^2,xy}_{\mathbf{R}_n} & \Psi^{xy}_{\mathbf{R}_n} & \Psi^{x^2-y^2,xy}_{\mathbf{R}_n}\\ 
\Psi^{z^2,x^2-y^2}_{\mathbf{R}_n} & \Psi^{xy,x^2-y^2}_{\mathbf{R}_n} & \Psi^{x^2-y^2}_{\mathbf{R}_n} 
\end{bmatrix},
\label{eq:nn_block}
\end{equation}  
where $\mathbf{R}_n=\mathbf{j}_n-\mathbf{i}$ is the vector connecting nearest-neighbor bonds for some fixed site $\mathbf{i}$, and $n\in[1,6]$ labels the six nearest-neighbors, and the spin indices for the matrix elements were again omitted.
There is a $3\times 3$ matrix for each direction $n$.
We find that all elements of $[\Psi_{{\mathbf{R}_n},\uparrow\downarrow}]$ are populated. 
All elements have a real and imaginary part,
and by comparing $[\Psi_{{\mathbf{R}_n},\uparrow\downarrow}]$ with $[\Psi_{{\mathbf{R}_n},\downarrow\uparrow}]$, we can identify $\mathrm{Re}\,[\Psi_{{\mathbf{R}_n},\uparrow\downarrow}]$ with the spin-singlet pairing correlations, and $\mathrm{Im}\,[\Psi_{{\mathbf{R}_n},\uparrow\downarrow}]$ with the spin-triplet. 
To make a connection with the paramterization introduced in \eqref{eq:gap_matrix}, we can write each element of \eqref{eq:nn_block} as 
\begin{equation}
\Psi^{\mu\nu}_{\mathbf{R}_n}=\mathrm{Re}\,\Psi^{\mu\nu}_{\mathbf{R}_n}+i\mathrm{Im}\,\Psi^{\mu\nu}_{\mathbf{R}_n}
= \psi^{\mu\nu}_{\mathbf{R}_n}+d^{\mu\nu}_{\mathbf{R}_n,z}. 
\end{equation} 
Among the matrix elements of \eqref{eq:nn_block},
the simplest and dominant is $\Psi^{z^2}_{\mathbf{R}_n}$, which corresponds to the rotationally symmetric orbital $d_{z^2}$. We, therefore, use $\Psi^{z^2}_{\mathbf{R}_n}$ as an illustrative example for the purpose of discussion. Whereas $\Psi_{z^2}$ in \eqref{eq:onsite_block} is necessarily singlet (see \eqref{singlets_no_SOC1}), $\Psi^{z^2}_{\mathbf{R}_n}$ is mixed. 
The self-consistent procedure reveals that
the correlations $\Psi^{z^2}_{\mathbf{R}_n}$ have a direction ($n$) dependent modulation of the form 
\begin{equation}
\Psi^{z^2}_{\mathbf{R}_n}=\left\langle c_{\mathbf{i}z^2\uparrow}c_{\mathbf{i}+\mathbf{R}_n,z^2\downarrow}\right\rangle =
\overbrace{\psi_\mathrm{nn}}^{\mathrm{singlet}}\overbrace{+i(-1)^{n+1}|d_{\mathrm{nn},z}|}^{\mathrm{triplet}},
\label{eq:z2pairing}
\end{equation}
The singlet component remains direction independent, but the imaginary triplet component induced by SOC alternates its sign from neighbor to neighbor, see figure \ref{fig:correlations}(c).
SOC induces the triplet component $|d_{\mathrm{nn},z}|$. 
Triplets are only suppressed through the coupling with the singlets. Because of this, at high magnetic fields, the triplets are favored. In fact, triplet Cooper pairs continue to condense with increasing magnetic field.
The relative amplitude $\varphi=\arctan(|d_{\mathrm{nn},z}|/\psi_\mathrm{nn})$ measures the increasing amount of the triplets over the singlets with increasing field. 
 
To gain more insight into $\Psi^{z^2}_{\mathbf{R}_n}$, it is instructive to examine the Fourier transform of \eqref{eq:z2pairing}. We obtain
\begin{align}
\label{eq:singtrip}
&\Psi^{z^2}_\mathrm{nn}(\mathbf{k}) =
\sum_{\mathbf{R}_n}\Psi^{z^2}_{\mathbf{R}_n}e^{i\mathbf{k}\cdot\mathbf{R}_n}=
\\ \nonumber 
&+2\psi_\mathrm{nn}\left[\cos k_x+2\cos\left(\frac{k_x}{2}\right)\cos\left(\frac{k_y\sqrt{3}}{2}\right)\right]\\ \nonumber 
& +4|d_{\mathrm{nn},z}|\left[-\sin k_x+2\sin\left(\frac{k_x}{2}\right)\cos\left(\frac{k_y\sqrt{3}}{2}\right)\right].
\end{align}
Both the singlet and triplet parts have nodal lines at which $\Psi^{z^2}_\mathrm{nn}(\mathbf{k})=0$. For the singlet component, the nodal line is closed is even in $\mathbf{k}$. For the triplet component, the nodal line coincides with the six $\Gamma M$ lines and is odd in in $\mathbf{k}$. The shape of the nodal line of $\Psi^{z^2}_\mathrm{nn}(\mathbf{k})=0$ evolves with increasing $\varphi$. At the critical angle $\varphi_c\approx 30^\circ$, the nodal line changes its topology, which is shown in figure \ref{fig:correlations}(d). Also, it is interesting to note that the momentum structure of the triplet component is identical to the SOC $g$-vector discussed in the context of Ising superconductivity in TMD's \cite{Youn2012,Xi2015,Saito2016,Nakamura2017,Nakata2018}; see appendix \ref{app:nodes} for more details. This is no coincidence and reflects the fact that triplets are induced by SOC. 

We stress that although \eqref{eq:singtrip} has a momentum dependence in the first Brillouin zone, it is in orbital basis, not band basis. This means that although \eqref{eq:singtrip} is nodal, this does not necessarily mean that the superconducting gap function in band basis is nodal. To obtain the band-dependent gaps from BdG, one would have to include the pairing correlations of all pairs on the lattice and rotate the Hamiltonian to band basis. We perform this analysis for a simplified model in appendix \ref{app:nodes}. 

The pairing correlations $\Psi^{xy}_{\mathbf{R}_n}$ and $\Psi^{x^2-y^2}_{\mathbf{R}_n}$ have a similar triplet modulation as $\Psi^{z^2}_{\mathbf{R}_n}$, but because of the lower symmetry of the orbitals, $|\Psi^{xy(x^2-y^2)}_{\mathbf{R}_n}|$ is direction dependent.   
For $\Psi^{x^2-y^2,xy}_{\mathbf{R}_n}$, it is the singlet component, not the triplet, that has a sign alternation. For the inter-orbital terms involving $z^2$, $\Psi^{xy(x^2-y^2),z^2}_{\mathbf{R}_n}$, both singlet and triplet component have a modulated sign. For the inter-orbital matrix elements, the symmetry $\Psi^{\alpha\beta}_{\mathbf{R}_n}=(\Psi^{\beta\alpha}_{\mathbf{R}_{n+3}})^*$ holds.

\subsubsection{On-site time-reversal-breaking correlations}  

A finite magnetic field induces non-unitary triplets with spin polarization $\mathbf{S}_I=i\langle \mathbf{d}_I^*\times\mathbf{d}_I\rangle$ and populate $[\Psi_{\mathbf{ii},\uparrow\uparrow}]$ and $[\Psi_{\mathbf{ii},\downarrow\downarrow}]$ with polarization direction pointing along the external magnetic in-plane field. For an in-plane field, the self-consistent procedure reveals a $d$-vector of the form $\mathbf{d}=(d_{I,x},d_{I,y},d_{I,z})$ and  $\mathbf{d}_I^*=(d_{I,x},d_{I,y},d_{I,z}^*)$, such that $d_{I,x}$ and $d_{I,y}$ are real, and $d_{I,z}$ is purely imaginary. The reality of $d_{I,x}$ and $d_{I,y}$ breaks time-reversal. 
Consequently, the components of polarization are
\begin{equation}
\begin{split}
\mathbf{S}_I& =i\left[d_{I,y}\left(d_{I,z}-d_{I,z}^*\right),-d_{I,x}\left(d_{I,z}-d_{I,z}^*\right),0\right]\\
&= 2\,\mathrm{Im}\,d_{I,z}\left(-d_{I,y},d_{I,x},0\right),
\end{split}
\end{equation} 
revealing no out-of-plane Cooper pair spin polarization, consistent with the in-plane magnetic field. 
Looking at on-site correlations and a magnetic field applied along $B_x$, $[\Psi_{\mathbf{ii},\uparrow\uparrow}]=[\Psi_{\mathbf{ii},\downarrow\downarrow}]=
d_{{\mathbf{ii}},y}^{x^2-y^2,xy}L_z/2$, where $L_z$ is given by \eqref{eq:half}, and $d_{\mathbf{ii},y}^{x^2-y^2,xy}$ is the $y$ component of the $d$-vector involving orbitals $4d_{x^2-y^2}$ and $4d_{xy}$. Since $d_{I,x}=0$ for on-site pairing correlations, $\mathbf{S}$ points along the $x$ direction as expected.

\subsection{Scalar impurities}
 
\begin{figure*}
\centering
\includegraphics[width=\textwidth]{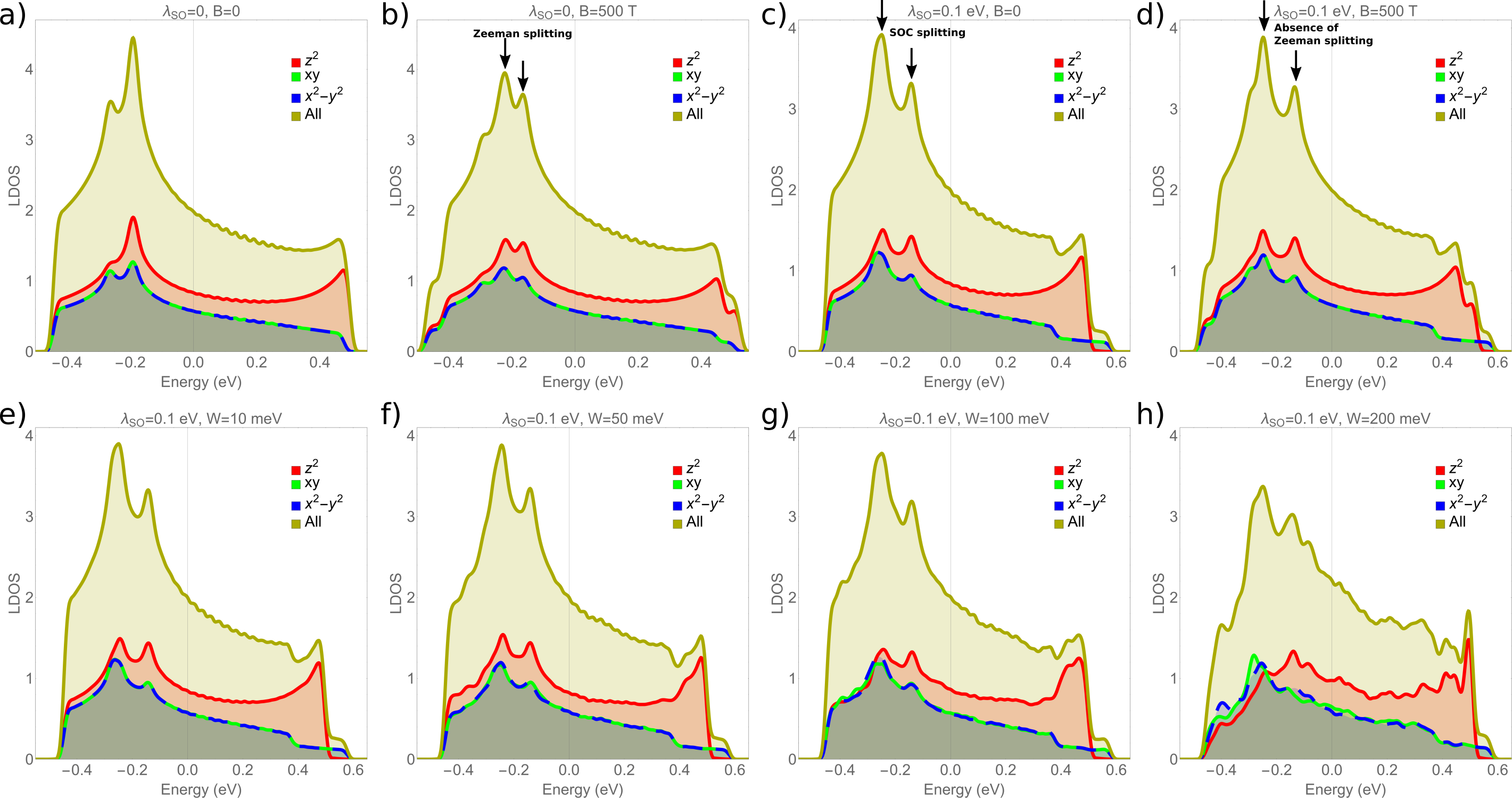}
\caption{\label{fig:disorder} 
The LDOS in the normal state for the Fermi level crossing bands for various combinations of SOC, in-plane magnetic fields, and Anderson disorder strengths. Here we simulated a $120\times 120$ triangular lattice with $N=1200$ Chebyshev expansion moments. The Gibbs oscillations are slightly visible. 
a) No Ising SOC at zero fields. 
b) Addition of an in-plane field of $B=500$ T. 
Zeeman spin-spitting is best seen at the top peaks, indicated by the arrows.
c) With Ising SOC at zero fields. The SOC magnetic induction acts as an effective out-of-plane Zeeman field but preserves time-reversal symmetry. The SOC-split bands are indicated by the arrows. 
d) Addition of a field of $B=500$ T, demonstrating the absence of Zeeman splitting due to the Ising locking mechanism. 
e) Same situation as in (c), but now with random on-site disorder strength $W=10$ meV, comparable to the superconducting energy scale. The probability distribution used here is uniform.
There is no visible difference (at the $E_\mathrm{F}$ scale) with respect to (c). 
f-h) Increasing $W$ up to $200$ meV.
Deviations between the partial LDOS correspondent to the $d_{xy}$ and $d_{x^2-y^2}$ (which are identical in the clean case) orbitals become more pronounced. 
}
\end{figure*}

In this section, we show how scalar on-site disorder affects the normal and the superconducting state. 

\subsubsection{LDOS of the normal state}

One of the hallmarks of Ising superconductors is the absence of Zeeman splitting due to the spin-locking by SOC \cite{Dvir2017}. 
To show this, we plot the local density of states (LDOS) $\rho_\mathbf{i}(E)$ \eqref{eq:ldos} with it's partial orbital contributions of an arbitrarily chosen atom with $\lambda_\mathrm{SO}=0$, and see a clear Zeeman splitting with the application of an in-plane field of $B=500$ T, where the splitting is $\Delta_\mathrm{Z}=\mu_\mathrm{B}B\approx 30$ meV; see figures \ref{fig:disorder}(a)-(b).     
Turning now to figure \ref{fig:disorder}(c), where $B=0$ with $\lambda_\mathrm{SO}=0.1$ eV, it seems as if the top peak is split by a Zeeman field, but actually corresponds to SOC splitting. For this reason, Ising SOC is frequently referred to as an effective Zeeman field, but one has to keep in mind that SOC preserves time-reversal symmetry, whereas a Zeeman field breaks it. 
For $\lambda_\mathrm{SO}=0.1$ eV, one can estimate the spin-orbit magnetic induction $B_\mathrm{SO}\approx \lambda_\mathrm{SO}/(2\mu_\mathrm{B})\approx 864$ T. Adding now an in-plane field of $B=500$ T (figure \ref{fig:disorder}d), no Zeeman splitting of the peaks appears. This explains the absence of Zeeman splitting in experiments \cite{Dvir2017}.  

In figures \ref{fig:disorder}(e)-(h) we show the case for finite $\lambda_\mathrm{SO}$ and $B=0$, for increasing uniformly distributed disorder with strength $W$. Usually, a disorder energy scale larger than the energy scale associated to unconventional superconducting energies $W>\Delta$ is strongly detrimental for unconventional superconducting states \cite{Mackenzie2003,Balatsky2006,Michaeli2012}. An appreciable difference of the partial LDOS contributed by the $d_{xy}$ and $d_{x^2-y^2}$ orbitals is only seen above a disorder strength of $W=100$ meV, that is, when the energy scale of disorder becomes comparable to $E_\mathrm{F}$.

\subsubsection{Robust unconventional superconductivity}

We now examine the effect of disorder on the superconducting correlations in \eqref{eq:onsite_block} and \eqref{eq:nn_block}. Usually, isotropic $s$-wave singlet superconducting correlations are robust against scalar impurities \cite{Anderson1959}, whereas unconventional correlations are strongly suppressed \cite{Balatsky2006}.

The Hamiltonian contains a randomly distributed on-site potential $W_{\mathbf{i}}^\mu$ with an orbital degree of freedom. The LDOS and the local correlations $[\Psi_{\mathbf{ii},\uparrow\downarrow}]$ vary from atom to atom. In principle, one now should do self-consistency for all $[\Psi_{\mathbf{ii},\uparrow\downarrow}]$ separately since they respond to the LDOS. However, we are interested in the overall behavior of the pairing correlations. For this reason, using an averaged value of $[\Psi_{\mathbf{ii},\uparrow\downarrow}]$ for all sites suffices for our purpose.

\begin{figure*}
\centering
\includegraphics[width=0.98\textwidth]{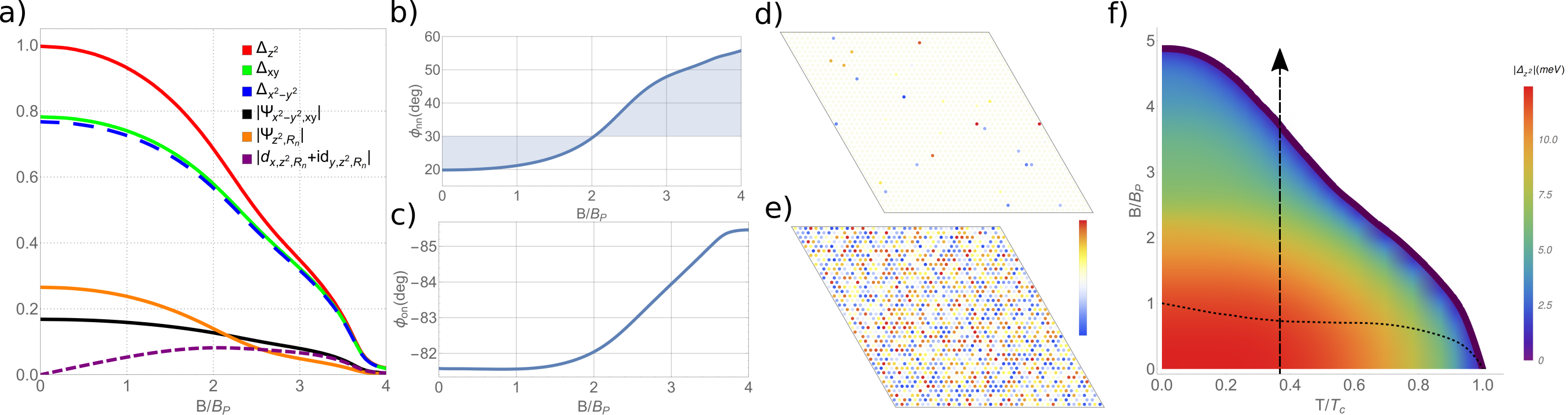}
\caption{\label{fig:robustness}
The effect of scalar impurities on the superconducting state. The qualitative properties as compared to the clean case remain unaltered.
(a) Superconducting order parameters and induced pairing correlations in the presence of disorder with Anderson disorder strength $W=100$ meV at $T=0.37T_c$. 
(b) The evolution of $\varphi=\arctan(|d_{\mathrm{nn},z}|/\psi_\mathrm{nn})$, and (c) $\varphi_\mathrm{on}=\arctan(\mathrm{Im}\,\Psi_{x^2-y^2,xy}/\mathrm{Re}\,\Psi_{x^2-y^2,xy})$. In the clean limit $\varphi_\mathrm{on}=\pm\pi/2$. Note the inverted axis to emphasize the relative increase of triplets with increasing field.
(d) Illustration of dilute scalar disorder at $C_\mathrm{imp}=2\%$ concentration and $W=1.8$ eV, and (e) random Anderson disorder on all sites. The color scale shows an arbitrary amplitude and serves as a guide to the eye. 
(f) Phase diagram for $W=100$ meV (uniform distribution) normalized with respect to the same values as in figure \ref{fig:diagrams}. The phase diagram for the dilute disorder is indistinguishable.}
\end{figure*}

In figure \ref{fig:robustness} we show the results of a simulation for a single uniformly distributed disorder realization with $W=100$ meV, an order of magnitude larger than the superconducting energy scale. Remarkably, no signs of suppression are observed.
Even a large amount of disorder does not affect the qualitative features present in the clean case. We only start seeing substantial suppression once $W\approx\lambda_\mathrm{SO}$.
In figure \ref{fig:disorder}(a),
one can see a slight difference in $\Psi_{xy}\neq \Psi_{x^2-y^2}$ due to the breaking of translational and rotational invariance. 
In contrast to the clean case, the on-site triplet $\Psi_{x^2-y^2,xy}=\mathrm{Re}\,\Psi_{x^2-y^2,xy}+i\mathrm{Im}\,\Psi_{x^2-y^2,xy}$ now acquires a small randomised real part. 
It therefore now has a phase evolution $\varphi_\mathrm{on}=\arctan(\mathrm{Im}\Psi_{x^2-y^2,xy}/\mathrm{Re}\Psi_{x^2-y^2,xy})$ with varying magnetic field, which in the clean case remained strictly constant with $\varphi_\mathrm{on}=\pm\pi/2$. The evolution of $\varphi_\mathrm{on}$ is shown in figure \ref{fig:robustness}(c). 
The most important qualitative features remain intact: a strong enhancement of the critical paramagnetic field and an increasing imbalance of the singlet and triplet components.

We also simulated dilute disorder with an impurity concentration of $2\%$, but very large disorder strength $W=1.8$ eV. Using a Gaussian probability distribution for the disorder potentials, this corresponds to an estimated dimensionless scattering rate \eqref{eq:scatteringrate} of $\hbar/(\tau k_\mathrm{B}T_c)\approx 20$, whereas superconducting energy scales are of order $\hbar/(\tau k_\mathrm{B}T_c)\sim 1$. The results are indistinguishable from the ones presented in figure \ref{fig:robustness}, showing that superconductivity remains robust regardless if it is strong dilute or Anderson disorder. 

The $A_2'$ $s$-wave triplets $\Psi_{x^2-y^2,xy}$ are on-site and do not depend on the momentum $\mathbf{k}$. For this reason, we checked if $\Psi_{x^2-y^2,xy}$ is intrinsically robust against the disorder, just like the conventional $A_1'$ $s$-wave singlets. To do this, we considered an attractive interaction in the triplet $A_2'$ channel only, that is, $V\neq 0$ in \eqref{H_S_inv} and all other channels set to zero. The $A_2'$ channel has an associated triplet order parameter $\Delta_{x^2-y^2,xy}=V\Psi_{x^2-y^2,xy}$ for which self-consistency is performed. Unlike $A_1'$, the $A_2'$ channel is not paramagnetically limited. In the clean case ($W=0$), we observe some variation of $\Delta_{x^2-y^2,xy}$ with the field due to electronic structure changes. The situation remains qualitatively the same with the addition of disorder. 
The $A_2'$ $s$-wave triplet state is therefore intrinsically robust against both paramagnetic limiting and scalar on-site impurities.

\section{Discussion \label{sec:discussion}} 

This section is dedicated to discussing the key implications of our results that add to the understanding of unconventional Ising superconductivity. Specifically, we focus on: the second-order paramagnetically limited phase transition, the protection that SOC grants to the superconducting state against both magnetic fields and on-site disorder, the role and structure of the induced superconducting triplet components, nodal topological superconductivity, and we contrast our results with existing literature and briefly discuss the role of Rashba SOC. 

\subsection{Second-order paramagnetic transition}

In conventional superconducting thin films such as aluminum and beryllium under applied in-plane magnetic fields, the paramagnetic effect determines the (upper) critical field  $B_\mathrm{c}\approx B_{c2}$.
This is because the orbital critical field is suppressed as the out-of-plane superconducting coherence length does not fit the monolayer \cite{Meservey1975,Meservey1994,Adams1998,Zocco2013,Khestanova2018}. The critical field $B_\mathrm{c}$ is then obtained by comparing the superconducting condensation energy $\rho_0\Delta_0^2$ with the normal state $(\chi_\mathrm{N}-\chi_\mathrm{S}) B^2_\mathrm{c}/\mu_0$ to obtain at $T=0$  \cite{Clogston1962,Matsuda2007,Bauer2012}
\begin{equation}
    B^2_\mathrm{c}(0)= \frac{\rho_0\Delta_0^2}{\mu_0(\chi_\mathrm{N}-\chi_\mathrm{S})}\overset{\chi_\mathrm{S}\rightarrow 0}{=} \frac{\Delta_0^2}{2\mu_\mathrm{B}^2},
\end{equation}
where the Pauli normal state susceptibility $\chi_\mathrm{N}\approx 2\mu_0\mu_\mathrm{B}^2\rho_0$, $\rho_0$ is the DOS at the Fermi level (per unit volume), and $\chi_\mathrm{S}$ is the magnetic susceptibility in the superconducting state. In singlet superconductors  $\chi_\mathrm{S}=0$, and because of the discontinuous difference between $\chi_\mathrm{N}$ and $\chi_\mathrm{S}$, the phase transition is of the first-order. In the present case, the non-unitary triplet Cooper pairs have an in-plane spin-polarization and give a finite contribution to the superconducting susceptibility $\chi_\mathrm{S}$. This causes the phase transition to be of second-order \cite{DelaBarrera2018,Sohn2018}. The enhancement of $B_\mathrm{c}$ occurs as $\chi_\mathrm{S}\rightarrow\chi_\mathrm{N}$.

\subsection{Disorder robust superconductivity}

The presence of SOC in the non-centrosymmetric crystal induces triplet pairing correlations leading to an unconventional parity-mixed superconducting state. Two natural questions regarding the effect of disorder arise: (1) is the unconventional state robust? (2) is the critical paramagnetic field suppressed?

\subsubsection{Stability of \texorpdfstring{$T_c$}{Tc} to the disorder}

Generally, the $T_c$ in the unconventional superconductors is suppressed by the disorder.
In our system, the superconducting state has a mixed parity.
It is, therefore, a priory not guaranteed to be stable against disorder.
The triplet $s$-wave superconductivity, in fact, is suppressed by non-magnetic disorder \cite{Asano2018}.
To examine the possible implication of this result in the present context, let us look more closely into the structure of the order parameter at the  $K$($K'$) corners of the Brillouin zone.
There are two non-zero Cooper correlations below $T_c$ at the $K$($K'$) point, $\psi_{K;+\uparrow} \psi_{K';-\downarrow}$ and $\psi_{K;+\downarrow} \psi_{K';-\uparrow}$; see equation \eqref{OP_list2} for details. Here, all the $\psi_{K(K');\pm\uparrow (\downarrow)}$ are annihilation operators of the Bloch electrons
at momenta $K$($K'$) in the orbital state, $|d_{x^2-y^2}\rangle \pm i |d_{xy}\rangle$ with spin $\uparrow(\downarrow)$. The other symmetry allowed combinations have an energy far above the Fermi level, and therefore play no role in superconductivity.

The triplet correlations $ \langle \hat{\Psi}_{\mathrm{triplet}} \rangle = \frac{1}{2} [\langle \psi_{K;+\uparrow} \psi_{K';-\downarrow} \rangle - \langle \psi_{K';-\uparrow} \psi_{K;+\downarrow} \rangle ]$ coexist with the singlet correlations $ \langle \hat{\Psi}_{\mathrm{singlet}} \rangle = \frac{1}{2} [\langle \psi_{K;+\uparrow} \psi_{K';-\downarrow}\rangle + \langle \psi_{K';-\uparrow} \psi_{K;-\downarrow} \rangle ]$ thanks to the SOC. 
In the case of pure triplet correlations, the two order parameters differ in sign,
$\langle \psi_{K;+\uparrow} \psi_{K';-\downarrow} \rangle = - \langle \psi_{K';-\uparrow} \psi_{K;+\downarrow} \rangle$.
Hence, following the argument of \cite{Asano2018}, the spin conserving inter-valley scattering is pair-breaking.
Indeed, as the orbital wave functions of paired electrons are switched ($(K+) \leftrightarrow (K'-)$) as a result of the impurity scattering, the triplet order parameter changes sign.
In our system, in addition to the inter-valley scattering, there are other similar sources of pair-breaking.
The gap near the $\Gamma$ differs from the gap near $K(K')$.
Quite generically the gap variations over the Fermi surface make the inter-band scattering pair-breaking \cite{Golubov1997}.

Notwithstanding the above arguments, our numerical results show that $T_c$ is stable against the disorder. We have no suppression of the $T_c$ even for a relatively strong disorder with the scattering rate exceeding the superconducting gap.
The resolution to this apparent contradiction lies in the suppression of the pair-breaking impurity scattering thanks to the orthogonality of the orbital wave-functions.
Specifically, the important bands are centered around $\Gamma$, $K$ and $K'$, see figure \ref{fig:bands}.
These bands receive weights predominantly from $|d_{z^2}\rangle$, $|d_{x^2-y^2}\rangle + i |d_{xy}\rangle$ and $|d_{x^2-y^2}\rangle - i |d_{xy}\rangle$ orbitals, respectively.
For all three points $\Gamma$, $K$ and $K'$, the rotation by $2\pi/3$ ($C_3$) around the $z$-axis is a symmetry operation.
It remains an approximate symmetry in the proximity of each of the above three symmetry points.
As the orbitals, $|d_{z^2}\rangle$, $|d_{x^2-y^2}\rangle + i |d_{xy}\rangle$ and $|d_{x^2-y^2}\rangle - i |d_{xy}\rangle$ acquire a factor of $1$, $e^{-i 2 \pi/3}$ and $e^{i 2 \pi/3}$ upon $C_3$ rotation respectively, the Bloch states at $\Gamma$, $K$ and $K'$ all transform differently under $C_3$.
It follows that any disorder potential, commuting with  $C_3$, does not induce scattering between the $\Gamma$, $K$ and $K'$ points, because all the three points are eigenstates of $C_3$ with different eigenvalues.
For instance, $0 = \langle K | [\mathcal{H}_D, C_3] | K'\rangle =( e^{i 2 \pi/3} - e^{-i 2 \pi/3} ) \langle K |\mathcal{H}_D | K'\rangle$ and as a result we have, $\langle K |\mathcal{H}_D | K'\rangle =0$. 
In our model, the on-site potential is certainly $C_3$ symmetric, which explains the observed stability of $T_c$ against a non-magnetic disorder.

Our arguments parallel the explanation of the stability of $T_c$ against the disorder in MgB$_2$ \cite{Mazin2002}.
This superconductor has two distinct bands relevant for the superconductivity with $\pi$ and $\sigma$ orbital character, respectively.
The suppressed scattering between the $\pi$ and $\sigma$ bands reconciles the strong enhancement of the resistivity with no change in $T_c$ as the non-magnetic disorder is added.

The $C_3$ symmetry plays a similar role of the approximate chiral symmetry stabilising the odd-parity superconductivity as discussed in \cite{Michaeli2012}.
In both cases, the possible paired states transform differently under a given symmetry. Then, the disorder respecting this symmetry does not cause the pair breaking.

\subsubsection{Effect of the disorder on \texorpdfstring{$B_c$}{Bc}} 

We now discuss the influence of the disorder on the critical field, $B_{c}$.
Let us first describe the situation in the clean case.
Without SOC, paramagnetic limiting destroys superconductivity when the Zeeman splitting $\Delta_\mathrm{Z}$ compares to $k_\mathrm{B} T_c$ \cite{Maki1964}. 
This suppression occurs as the states with opposite spin polarization and opposite momenta differ by Zeeman splitting in energy. As a result, the Cooper logarithms take the form, $\log(\Lambda/ \Delta_\mathrm{Z})$ instead of $\log(\Lambda/ (k_\mathrm{B} T))$, where $\Lambda$ is the ultra-violet energy cut off of the order of the Debye energy, therefore the $T_c(B) < T_c = T_c(B=0)$.
One may consider the paring of the states with the same spin polarization. Such pairs, however, would necessarily be a spin triplet. In the absence of the attraction in spin triplet channel, such a pairing cannot be realized. 

In the presence of SOC and in-plane magnetic field, the residual symmetry $\mathcal{T}_h=\sigma_h\mathcal{T}$ ensures that each state of a given momentum $\mathbf{k}$ ,$|\phi_{\mathbf{k}}\rangle$, is degenerate with the state with the opposite momentum, $|\phi_{-\mathbf{k}}\rangle= \mathcal{T}_h |\phi_{\mathbf{k}}\rangle$, see figure \ref{fig:tilt}. 
This degeneracy ensures that the Cooper logarithms constructed on these states, $\log(\Lambda/ (k_\mathrm{B}T))$ are not suppressed.
Crucially, in contrast to the case without SOC, the degenerate pairs have a finite amplitude $\lambda_{\mathrm{SO}}/ \sqrt{\lambda_{\mathrm{SO}}^2 + \Delta_\mathrm{Z}^2}$ to be in the spin singlet state and enjoy the attraction.
Ignoring for the sake of the argument the $d_{z^2}$ orbital, we conclude that the original attraction  $U$ is renormalised as $U' = U\lambda^2_{\mathrm{SO}}/ (\lambda_{\mathrm{SO}}^2 + \Delta_\mathrm{Z}^2)$ \cite{Sosenko2017}.
This leads to the Gaussian dependence of the critical temperature on the in-plane field, $T_c(B) = T_c \exp[ -B^2 /B_0^2 ]$, where $g \mu_B B_0 \propto  \lambda_{\mathrm{SO}} [U \rho(E_\mathrm{F})]$.
Alternatively, it is consistent with the inverted Gaussian shape of the critical field temperature dependence, $B_{c}(T)$, with its characteristic inflexion point, see figure \ref{fig:robustness}(f).
At low temperatures we have $B_c(T) \propto \sqrt{\log(T_c/T)}$ in agreement with \cite{Ilic2017}.
The actual transition line is rounded at small temperatures, see figure \ref{fig:robustness}(f) as the superconducting gap becomes small at large magnetic fields.

From the picture presented above, it is clear that the impurities may affect $B_c$ only via the inter-band and/or inter-valley scattering.
Again, similar to the discussion in the previous section, as such scattering is prohibited by $C_3$ symmetry, no suppression of $B_c$ by impurities is expected.
This is indeed what we have found numerically.

\subsection{Nodal superconducting phase\label{sec:disc_nodal}}

\begin{figure*}
\includegraphics[width=0.95\textwidth]{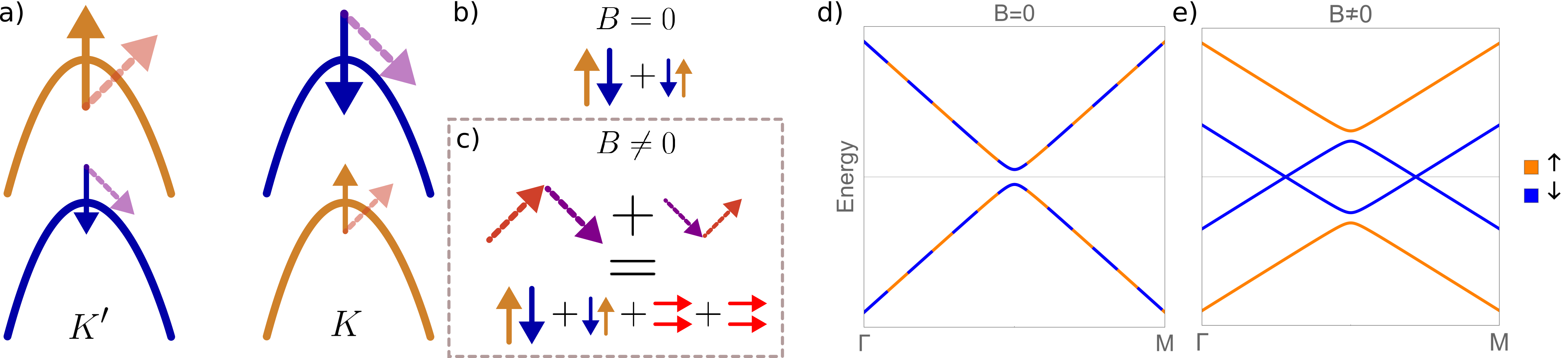}
\caption{\label{fig:tilt}
a) Schematic illustration of the adjustment of the Cooper pairs to an in-plane magnetic field. The size of the arrows reflect the amplitude of the order parameters. The Cooper spin partners $\nearrow$ and $\searrow$ are related by the symmetry $\mathcal{T}_h=\sigma_h\mathcal{T}$, which ensure their degeneracy at opposite momenta in 2D. Schematically, $\sigma_h\mathcal{T}\nearrow=\sigma_h\swarrow=\searrow$. Cooper partners are paired at the same energy located at $K$ and $K^\prime=\mathcal{T}K$. b) At $B=0$, Cooper pairs respect time-reversal symmetry. c) A finite $B$ induces non-unitary triplets breaking time-reversal (red arrows). d) Schematic quasi-particle dispersion along $\Gamma M$, where SOC vanishes. At zero field, the bands are spin-degenerate, opening a pure singlet gap around the Fermi level. e) At a finite field in the range $B_\mathrm{T}<B<B_\mathrm{c}$, a pair of nodes form along $\Gamma M$. Because the spin-degeneracy around the Fermi level is now lifted, no singlet gap can open. These nodes occur strictly along $\Gamma M$. If one slightly deviates from the $\Gamma M$ line, a finite SOC together with the Zeeman field induce non-unitary equal spin Cooper pairs, opening up a pure triplet gap around the Fermi level.}
\end{figure*}

The possibility of driving a TMD monolayer to a nodal topological superconductor supporting Majorana fermions has recently been considered \cite{He2018,Fischer2018}.  
Here we argue that the quasi-particle dispersion is inevitably nodal at a high magnetic field, and they occur strictly on $\Gamma M$.
The fact that SOC vanishes on $\Gamma M$ has an important effect on the quasi-particle dispersion at high magnetic fields. 
While SOC is finite in the quasi-particle spectrum region $E(\mathbf{k}\neq \Gamma M)$ and Zeeman splits very weakly, the lines $E(\mathbf{k}= \Gamma M)$ are strongly Zeeman split. Then, at high fields, $E(\mathbf{k}=\Gamma M)$ develops a pair of nodes along each $\Gamma M$ line; see figure \ref{fig:tilt}(e). At some critical magnetic field $B_\mathrm{T}$ in the range $B_\mathrm{P}\lesssim B_\mathrm{T} < B_\mathrm{c}$, the superconducting phase transitions from a fully gapped phase to a nodal phase. Here $B_\mathrm{P}$ is the paramagnetic limit in the absence of SOC, and $B_\mathrm{c}$ is the paramagnetic limit in the presence of SOC. The nodal transition is driven by the Zeeman effect, not by the intrinsic nodal structure of the superconducting triplet component in \eqref{eq:singtrip}. In appendix \ref{app:nodes} we provide an explicit model and demonstrate that the nodal phase appears at high fields.   

\subsection{Residual chiral symmetry and topology} 

We now comment on the topology of the nodal phase. Following the developments of references \cite{Bzdusek2017,Fischer2018}, the basal mirror symmetry $\sigma_h$ present in monolayer TMDs has important implications for superconductivity in 2D.  
In 3D, in the presence of both time reversal and inversion symmetries, the states at each momentum $\bf{k}$ are doubly degenerate. The two states at $\bf{k}$ and two states at $-\bf{k}$ can be combined to form one Cooper pair which is parity even-singlet, or three Cooper pairs which are parity odd-triplets \cite{Anderson1984}. In the absence of the above discrete symmetries, the simple classification of Cooper pairs does not apply, which generally affects the superconductivity.
In 2D, the additional symmetries $C_{2z}=\sigma_h\mathcal{I}$ and $\mathcal{T}_h=\sigma_h\mathcal{T}$ also guarantee degeneracy at opposite momenta. 
The Hamiltonian reported here lacks $C_{2z}$, but still has $\mathcal{T}_h$.

Within an extended Altland-Zirnbauer (AZ$+\mathcal{I}$) classification scheme \cite{Bzdusek2017,Fischer2018}, the gapped superconducting phase below $B_\mathrm{T}$ is in the BDI class without topological charges and no Majorana edge states. Above $B_\mathrm{T}$, the class transitions to a nodal AIII class with associated $\mathbb{Z}$ topological charges. 
The symmetries allowing for a topological classification of point nodes supporting Majorana flat bands are $\mathfrak{I}=C_{2z}\mathcal{T}_h$, $\mathfrak{B}=C_{2z}\mathcal{P}$, where $\mathcal{P}$ refers to particle-hole symmetry; and more importantly the chiral symmetry $\mathfrak{C}=\mathfrak{I}\mathfrak{B}$ \cite{Schnyder2011,Schnyder2012,Bzdusek2017,Fischer2018}.  
These operators fulfil 
\begin{align}
\mathfrak{I} \mathcal{H}(\mathbf{k})\mathfrak{I}^{-1} & =\mathcal{H}(\mathbf{k}), &\mathfrak{I}^2 =\pm 1& \quad (\mathrm{AU}), \nonumber \\
 \mathfrak{B}\mathcal{H}(\mathbf{k})\mathfrak{B}^{-1} & =-\mathcal{H}(\mathbf{k}), &\mathfrak{B}^2 =\pm 1& \quad (\mathrm{AU}),\\
 \mathfrak{C}\mathcal{H}(\mathbf{k})\mathfrak{C}^{-1} & =-\mathcal{H}(\mathbf{k}), &\mathfrak{C}^2 = 1& \quad (\mathrm{U}), \nonumber
\end{align} 
where $\mathcal{H}(\mathbf{k})$ is a BdG Hamiltonian in $\mathbf{k}$-space, AU indicates anti-unitarity and U unitarity. 
With only $\mathcal{T}_h$ present, monolayer TMDs lack $\mathfrak{I}$ and $\mathfrak{B}$, but still preserve the \textit{residual} chiral symmetry $\mathfrak{C}=\mathfrak{I}\mathfrak{B}=\mathcal{T}_h\mathcal{P}$, which provides protection for the topological point nodes.
The chiral symmetry $\mathfrak{C}$ is not affected by on-site scalar impurities \cite{Chiu2016}.

\subsection{The effect of Rashba SOC}

Experimental setups might also produce Rashba SOC coming from a substrate. In contrast to Ising SOC, Rashba SOC locks the spins in the in-plane direction with helical texture \cite{Gorkov2001,Bauer2012,Smidman2017}. 
This populates  $\Psi_{I,\uparrow\uparrow}$ and $\Psi_{I,\downarrow\downarrow}$ in \eqref{eq:gap_matrix}, which correspond to triplet Cooper pairs that are unprotected against in-plane magnetic fields. Therefore, the inclusion of Rashba SOC suppresses the paramagnetic critical field, as it competes for spins with Ising SOC \cite{Ugeda2015,DelaBarrera2018}. 

In this paper, we only discussed on-site scalar disorder, which preserves the $D_{3h}$ symmetry of the lattice. However, in experimental settings, the monolayer is interfaced with other materials via encapsulation or contact with a substrate. The lattice mismatch at the interfaces is expected to introduce long-range scattering potentials, which can break the $D_{3h}$ symmetry of the monolayer. In this case, we speculate that $C_3$-breaking long-range disorder might lead to suppression of both $T_c$ and $B_c$.

\section{Conclusion}

When little is known about the superconducting pairing mechanism, group theory traditionally allows one to lay out the menu of possible pairing symmetries, and study the most likely realizations \cite{Sigrist1991}. In many cases, it is possible to pinpoint the paring symmetries without detailed knowledge of the pairing interaction. Among the (sometimes) large menu of possible gap structures offered by group theory, the question of which ones are in fact realized remains generally open. 
In non-centrosymmetric materials, a parity-mixed superconducting state is allowed, but the degree in which singlets and triplets mix, remains mostly unclear on solely group theoretical grounds. 
In this work,
the self-consistent BdG theory provides us with the pairing amplitudes of the pairing correlations classified by group theory. By assuming a pairing interaction in the conventional $s$-wave $A_1'$ singlet channel only, we have shown that the resulting superconducting state is parity-mixed.  

Using this combined group theoretical and real-space numerical approach, we self-consistently obtained the unconventional superconducting state of 2D NbSe$_2$,
and investigated how it responds to an external in-plane magnetic field and scalar impurities. 
We focused on on-site and nearest-neighbor pairing correlations. Because of the orbital degree of freedom, an on-site $A_2'$ triplet is induced by SOC which is structure-less in $\mathbf{k}$-space and intrinsically robust against the disorder. The magnetic field increases the imbalance between triplet and singlet Cooper pairs and induces a non-unitary component breaking time-reversal. 

Ising spin-orbit coupling not only enhances the upper critical field but also ensures robustness of the parity-mixed superconducting state against the usually detrimental scalar impurities.
Moreover, the multi-orbital nature of monolayer TMDs is important for the robustness against the disorder.
The orthogonality of the orbital wave-functions prevents inter-band scattering. In this regard, taking into account the multi-orbital nature of TMDs essential, because it cannot be mapped to a minimal graphene model with two inequivalent sublattices.

Although the system lacks both time-reversal and inversion, monolayer TMDs subjected to in-plane magnetic fields possess a residual chiral symmetry $\mathfrak{C}$, which is a non-spatial symmetry providing topological protection for Majorana flat bands. Even though we did not elaborate on the details of the nodal topological phase (which is left for a future prospect), our results are consistent with recent reports of Zeeman field driven nodes in the quasi-particle spectrum on $\Gamma M$ \cite{He2018,Fischer2018}. 

We used monolayer NbSe$_2$ as our base system, but our work is also relevant for all TMD families. In fact, most of the TMDs are very similar in band-structure and orbital weights \cite{Liu2013}, differing essentially in the position of the chemical potential. For this reason, we speculate that strongly hole-doped group-6 TMD show potential for superconducting applications.

In the quest for new group-6 2D TMD superconductors, it might be interesting to investigate possible superconducting states in strongly hole-doped tungsten dichalcogenides such as WS$_2$, WSe$_2$ and WTe$_2$ \cite{Eftekhari2017}. The Tungsten family is the heaviest among the group-6 TMD and offers a giant spin-splitting \cite{Zhu2011}. Other advantages as compared to the Molybdenum family is that Tungsten is more abundant in nature, cheaper and less toxic. On the other hand, most group-5 TMDs are metals, such as NbSe$_2$ and TaS$_2$. From the SOC perspective, Tantalum based materials yield a larger spin-splitting.

\begin{acknowledgments}
We thank Girsh Blumberg, Tom Dvir, Zohar Ringel, Hadar Steinberg and Vilen Zevin for helpful and enlightening discussions.
We recognize the financial support by the Israel Science Foundation, Grant No. 1287/15. 
\end{acknowledgments}
 
\clearpage

\appendix

\section{Real-space tight-binding model \label{app:tb}}
 
\begin{figure}
\centering
\includegraphics[width=0.48\textwidth]{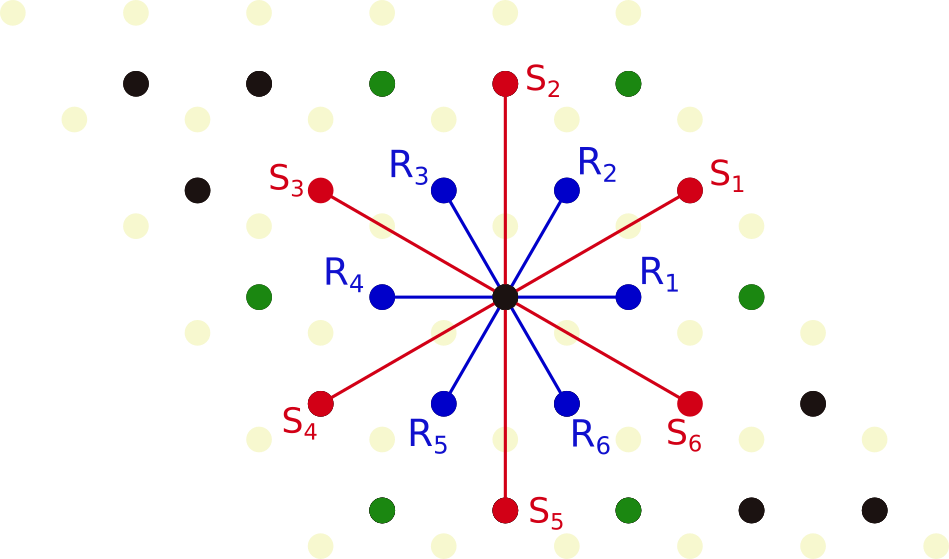}
\caption{\label{fig:neighbors} Illustration of the first (blue) $\mathbf{R}_n$, second (red) $\mathbf{S}_n=\mathbf{R}_n+\mathbf{R}_{n+1}$ and third nearest neighbors (green) $\mathbf{T}_n=2\mathbf{R}_n$ on the triangular lattice. Chalcogen atoms are shown by faint yellow points.   }
\end{figure}

In this section, we specify the tight-binding model introduced in \eqref{eq:h0}. This discussion applies to all TMD monolayers with hexagonal polytype 1H. Without SOC, the tight-binding model consists of the hoppings $t_{\mathbf{ij}}^{\mu\nu}$ on a transition metal triangular lattice. Hopping distance between atomic lattice sites $\mathbf{i}$ and $\mathbf{j}$ is included up to third nearest neighbors, and the orbital indices $\mu$ and $\nu$ run over the minimal set of orbitals $\{d_{z^2},d_{xy},d_{x^2-y^2}\}$. On the triangular lattice, they transform according to the point group of monolayer TMD's $D_{3h}$.
For each fixed direction of $\langle\mathbf{i},\mathbf{j}\rangle$, there is a corresponding $3\times 3$ matrix in orbital space specifying all the intra-orbital ($\mu=\nu$) and inter-orbital ($\mu\neq\nu$) hoppings. Considering the six nearest-neighbors, six second second-nearest-neighbors, six third-nearest neighbors, and three orbitals; we therefore have $(3\times 6)\times 3^2=162$ matrix elements $t_{\mathbf{ij}}^{\mu\nu}$ to specify. Using the values of the overlap integrals along the $t_{\mathbf{R}_1}^{\mu\nu}$ direction as reference to obtain all the others, we list the $162$ matrix elements in table \ref{tab:tbtable}.  Nearest-neighbor matrices are labelled by $\hat{R}_n$, second-nearest-neighbor by $\hat{S}_n$, and third-nearest-neighbor by $\hat{T}_n$, where $n$ runs from $1$ to $6$, as can also be seen in figure \ref{fig:neighbors}. 

The tight-binding parameters that fit the transition metal dominated bands in 1H-NbSe$_2$ were adapted from reference \cite{He2018}, and are listed in table \ref{tb:tbparameters}. The Fourier transform of our real-space tight-binding model in equation \eqref{eq:h0} with all the hoppings $t_{\mathbf{ij}}^{\mu\nu}$ given in table \ref{tab:tbtable} yields the $\mathbf{k}$-space tight-binding model developed by Liu \textit{et al} \cite{Liu2013}.
A good checkup for the matrices in table \ref{tab:tbtable} is the property 
\begin{equation}
\frac{1}{6}\sum_{n=1}^6\hat{R}_n=\begin{bmatrix}
t_0 & 0 & 0\\ 
0 & t_{11}+t_{12}/\sqrt{3} & 0\\ 
0 & 0 & t_{11}+t_{12}/\sqrt{3}
\end{bmatrix},
\label{eq:sumR}
\end{equation}
and analogously for $\hat{T}_n$; and for the second nearest-neighbor matrices
\begin{equation}
\frac{1}{6}\sum_{n=1}^6\hat{S}_n=\begin{bmatrix}
r_0 & 0 & 0\\ 
0 & (r_{11}+r_{22})/2 & 0\\ 
0 & 0 & (r_{11}+r_{22})/2
\end{bmatrix}.
\label{eq:sumS}
\end{equation}
The matrix elements for the in-plane orbitals are the same, which reflects the fact that they belong to the same irrep $E'$. Together with the pairing interaction \eqref{eq:interaction}, equations \eqref{eq:sumR} and \eqref{eq:sumS} enforce $\Delta_{x^2-y^2}=\Delta_{xy}$ for the clean case. 

The value of $\lambda_\mathrm{SO}$ that fits the band structure from first principle calculations is $\lambda_\mathrm{SO}=78.4$ meV. In figure \ref{fig:bands} we used $\lambda_\mathrm{SOC}=100$ meV for presentation, and the calculations were performed using $\lambda_\mathrm{SO}=200$ meV, to obtain a sizable ratio $\lambda_\mathrm{SO}/\Delta_{z^2}(0)$ as explained in section \ref{sec:results}. The value of $\lambda_\mathrm{SO}=200$ meV is still low enough to avoid that the chemical potential lies between the two spin-split bands crossing the Fermi level, which is the situation described in references \cite{Hsu2017,Sosenko2017}.

\begin{table}
\centering
\caption{Tight-binding fitting parameters for 1H-NbSe$_2$ adapted from reference \cite{He2018}. The units are in eV. 
The hoppings are overlap integrals defined along the $\mathbf{R}_1$ direction.
}
\label{tb:tbparameters}
\textbf{1H-NbSe$_2$}
\begin{ruledtabular}
\begin{tabular}{cccccc}
$t_0$ & $t_1$ & $t_2$ & $t_{11}$ & $t_{12}$ & $t_{22}$ \\
$-0.2308$ & $0.3116$ & $0.3459$ & $0.2795$ & $0.2787$ & $-0.0539$ \\
$r_0$ & $r_1$ & $r_2$ & $r_{11}$ & $r_{12}$ & $r_{22}$ \\
$0.0037$ & $-0.0997$ & $-r_1/\sqrt{3}$ & $0.0320$ & $0.0986$ & $0$ \\
$u_0$ & $u_1$ & $u_2$ & $u_{11}$ & $u_{12}$ & $u_{22}$ \\
$0.0685$ & $-0.0381$ & $0.0535$ & $0.0601$ & $-0.0179$ & $-0.0425$ \\
$\epsilon_0$ & $\epsilon_1$ & $\epsilon_2$ & $\mu_0$ & $\lambda_\mathrm{SO}$ &  \\
$1.4466$ & $1.8496$ & $1.8496$ & $0$ & $0.0784$ & 
\end{tabular}
\end{ruledtabular}
\end{table}

\section{Zeeman field driven nodes in the quasi-particle spectrum \label{app:nodes}}
 
The evolution of the quasi-particle spectrum with the magnetic field is peculiar along high symmetry lines where SOC vanishes. Here we argue that the quasi-particle spectrum develops nodes along the high symmetry line $\Gamma M$ where superconductivity remains purely singlet, which leads to a nodal superconducting phase at high fields.
The orbital degree of freedom and a specific band structure is unimportant for this discussion's sake, and for this reason, we consider a simpler model without the orbital degree of freedom. We stress that the nodes generated through the arguments presented here come from the Zeeman field, not from the triplet component. 

In a simple pseudo single-orbital picture, one can model the anti-symmetric spin-orbit term via \cite{Youn2012,Xi2015}
\begin{equation}
\begin{split}
\mathcal{H}_\mathrm{ASOC} &  =  -i\lambda\sum_{n,\sigma,\sigma^\prime}(\nabla V_n\times\mathbf{R}_n)\cdot\boldsymbol{\sigma}_{\sigma\sigma^\prime}c^\dag_{\mathbf{i}\sigma}c_{\mathbf{i}+\mathbf{R}_n,\sigma^\prime}\\
&
=\sum_{\mathbf{k},\sigma,\sigma^\prime}\mathbf{g}(\mathbf{k})\cdot\boldsymbol{\sigma}_{\sigma\sigma^\prime}c^\dag_{\mathbf{k}\sigma}c_{\mathbf{k}\sigma^\prime},
\end{split}
\label{eq:procedure}
\end{equation}
where 
\begin{equation}
\mathbf{g}(\mathbf{k})=\frac{\lambda}{2}\left[-\sin k_x+2\sin\left(\frac{k_x}{2}\right)\cos\left(k_y\frac{\sqrt{3}}{2}\right)\right]\boldsymbol{\hat{z}}.
\label{eq:isingSOC}
\end{equation}
Here $\nabla V_n$ is the crystal field unit vector direction according to the blue arrows in figure \ref{fig:crystal}b, $\mathbf{R}_n$ are the nearest neighbor vectors and the second line in the Fourier transformed version in momentum space with the typical form of the anti-symmetric $g$-vector. Equation \eqref{eq:isingSOC} is found in several references modelling Ising superconductors\cite{Youn2012,Xi2015,Saito2016,Nakamura2017,Nakata2018}, and the procedure  \eqref{eq:procedure} is a simple motivation to quickly obtain $\mathbf{g}(\mathbf{k})$. Uncoincidentally, \eqref{eq:isingSOC} has the same structure as the triplet component, as discussed below equation \eqref{eq:singtrip}. 

Therefore, we consider the Hamiltonian 
\begin{equation}
\begin{split}
\mathcal{H}(\mathbf{k}) = & \sum_{\mathbf{k},\sigma}\epsilon(\mathbf{k})c^\dag_{\mathbf{k}\sigma}c_{\mathbf{k}\sigma}+\sum_{\mathbf{k},\sigma\sigma^\prime} \mathbf{g}(\mathbf{k})\cdot\boldsymbol{\sigma}_{\sigma\sigma^\prime}\,c^\dag_{\mathbf{k}\sigma}c_{\mathbf{k}\sigma^\prime}\\
& -\mu_\mathrm{B}\sum_{\mathbf{k},\sigma\sigma^\prime} \mathbf{B}\cdot\boldsymbol{\sigma}_{\sigma\sigma^\prime}\,c^\dag_{\mathbf{k}\sigma}c_{\mathbf{k}\sigma^\prime}\\
&+\sum_{\mathbf{k},\sigma\sigma^\prime}\left[\Delta_{\sigma\sigma^\prime}(\mathbf{k})c^\dag_{\mathbf{k}\sigma}c^\dag_{-\mathbf{k}\sigma^\prime}+\mathrm{h.c.} \right ],
\end{split}
\label{eq:toy}
\end{equation} 
Here $\epsilon(\mathbf{k})=\epsilon(-\mathbf{k})$ is the symmetric part of the band-structure. The $g$-vector is given by \eqref{eq:isingSOC}. The magnetic field
$\mathbf{B}$ is the Zeeman magnetic induction, and $\Delta_{\sigma\sigma^\prime}(\mathbf{k})$ includes both singlet and triplet pairing.  
Without loss of generality, we consider an in-plane magnetic field $\mathbf{B}=B\hat{\mathbf{x}}$, and hence a $d$-vector of the form $\mathbf{d}(\mathbf{k})=(0,d_y(\mathbf{k}),d_z(\mathbf{k}))$. There is no $d_x$ component because we are limiting the in-plane field to the $x$ direction, and Cooper pair spin polarization points along $i\mathbf{d}\times\mathbf{d}^*\parallel \mathbf{B}$, which enforces $d_x=0$. 
Then,  $\Delta_{\uparrow\uparrow}(\mathbf{k})=\Delta_{\downarrow\downarrow}(\mathbf{k})=id_y(\mathbf{k})$, $\Delta_{\uparrow\downarrow}(\mathbf{k})=\psi(\mathbf{k})+d_z(\mathbf{k})$ and $\Delta_{\downarrow\uparrow}(\mathbf{k})=-\psi(\mathbf{k})+d_z(\mathbf{k})$. 

In the basis of $\{c^\dag_{\mathbf{k}\uparrow},c^\dag_{\mathbf{k}\downarrow},c_{-\mathbf{k}\uparrow},c_{-\mathbf{k}\downarrow}\}$, the Hamiltonian \eqref{eq:toy} can be written as a $4\times 4$ matrix $[\mathcal{H}(\mathbf{k})]$ given by ($\mu_\mathrm{B}=1$)
\begin{align}
 \label{eq:kronecker}
& [\mathcal{H}(\mathbf{k})] = \\   
&  {\scriptsize  
\begin{bmatrix}
\epsilon(\mathbf{k})+g(\mathbf{k}) & -B & id_y(\mathbf{k}) & \psi(\mathbf{k})+d_z(\mathbf{k})
\\ 
-B & \epsilon(\mathbf{k})-g(\mathbf{k}) & -\psi(\mathbf{k})+d_z(\mathbf{k}) 
& id_y(\mathbf{k})\\ 
-id^*_y(\mathbf{k}) & -\psi^*(\mathbf{k})+d_z^*(\mathbf{k}) 
& -\epsilon(\mathbf{k})+g(\mathbf{k}) & B\\
\psi^*(\mathbf{k})+d_z^*(\mathbf{k}) 
& -id^*_y(\mathbf{k}) & B & -\epsilon(\mathbf{k})-g(\mathbf{k})
\end{bmatrix}.
}%
\notag  
\end{align}  

We now analyse the quasi-particle dispersion $E(\mathbf{k})$ determined by the characteristic equation $\det([\mathcal{H}(\mathbf{k})]-E(\mathbf{k})\mathbb{I})=0$ in some detail looking at specific cases. 

\subsection{Zero magnetic field}

If $B=0$, the Hamiltonian must preserve time-reversal and hence $d_y(\mathbf{k})=0$ \footnote{Assuming that the superconducting state does not break time-reversal symmetry spontaneously.}.  
Then, the dispersion simplifies to a familiar BCS-like form
\begin{equation}
E^2(\mathbf{k})=\left(\epsilon(\mathbf{k})\pm g(\mathbf{k}) \right )^2+|\Delta_\pm(\mathbf{k})|^2,
\end{equation}
where
\begin{align}
|\Delta_\pm(\mathbf{k})|^2=& +|\psi(\mathbf{k})|^2+|d_z(\mathbf{k})|^2 \notag \\
& \pm\left(\psi(\mathbf{k})d_z^*(\mathbf{k})+\psi^*(\mathbf{k})d_z(\mathbf{k}) \right ).
\end{align} 
The Ising SOC $g(\mathbf{k})$ yields split bands, which have a superconducting gap $\Delta_\pm(\mathbf{k})$. The singlet-triplet coupling $\psi d_z^*+\psi^* d_z=2|\psi||d_z|\cos(\varphi_s-\varphi_t)$, where $\varphi_{s(t)}$ is the phase of the singlet (triplet) component, causes the spin-split bands to have different gap values, namely $\Delta_+$ and $\Delta_-$. Nodes in the quasi-particle spectrum $E(\mathbf{k})$ are only possible if $\Delta_\pm(\mathbf{k})$ itself is nodal. 

\subsection{No SOC}

If $\mathbf{g}(\mathbf{k})=\mathbf{d}(\mathbf{k})=0$, we obtain the situation of a Zeeman split singlet superconductor with dispersion
\begin{equation}
    |E(\mathbf{k})|=\left|\sqrt{\epsilon^2(\mathbf{k})+|\psi|^2}\pm B \right |.
    \label{eq:zeeman}
\end{equation}
All Cooper pairs are depaired once the Zeeman energy compares to the superconducting condensation energy.  No nodal superconducting phase arises. 

Alternatively, we also can look at $\mathbf{g}(\mathbf{k})=\psi(\mathbf{k})=0$. This corresponds to a purely triplet superconductor with dispersion
\begin{equation}
\begin{split}
E^2(\mathbf{k})=&+\left(\epsilon(\mathbf{k})\pm B\right)^2+|d_y(\mathbf{k})|^2+|d_z(\mathbf{k})|^2 \\
&\pm i\left(d_z(\mathbf{k})d^*_y(\mathbf{k})-d^*_z(\mathbf{k})d_y(\mathbf{k})\right).
\end{split}
\end{equation} 
The second line corresponds to the non-unitary triplet part $i\mathbf{d}\times\mathbf{d}^*=-i(d_z d_y^*-d_z^* d_y)=2|d_z||d_y|\sin(\varphi_z-\varphi_y)$ that breaks time-reversal. 
A magnetic field does not suppress the $d$-vector as opposed to the singlet case.

\subsection{With SOC}
 
We now include both SOC and magnetic field and set $d_y=0$. This is justified if the SOC magnetic induction is much larger than the external in-plane field. Then, we obtain
\begin{widetext}
\begin{equation}
E^2(\mathbf{k})=\epsilon^2(\mathbf{k})+g^2(\mathbf{k})+B^2+|\psi(\mathbf{k})|^2+|d_z(\mathbf{k})|^2\pm\sqrt{\left(2g(\mathbf{k})\epsilon(\mathbf{k})+\psi(\mathbf{k})d^*_z(\mathbf{k})+\psi^*(\mathbf{k})d_z(\mathbf{k}) \right )^2+4B^2\left(\epsilon^2(\mathbf{k})+|\psi(\mathbf{k})|^2 \right ) }.
\label{eq:disp}
\end{equation} 
\end{widetext}
SOC induces the mixed term  $\psi d_z^*+\psi^* d_z$.
The term $B^2|\psi|^2$ couples the Zeeman field to the singlet order parameter, and is responsible for paramagnetic limiting \cite{Mockli2018}. Note that there is no such term for the triplets since they do not suffer paramagnetic limiting. 
However, the cross term $\psi d_z^*+\psi^* d_z$ indirectly suppresses the triplet component.
The simultaneous presence of Ising SOC and the Zeeman field allows for the possibility of magnetic field driven nodes in the quasi-particle spectrum $E(\mathbf{k})$ along $\Gamma M$. 
Along $\Gamma M$ $g(\mathbf{k})=0$, and the dispersion \eqref{eq:disp} reduces to \eqref{eq:zeeman} and the absence of SOC eliminates triplet superconductivity along this line.
Therefore, at sufficiently high fields, the quasi-particle dispersion $E(\mathbf{k})$ develops a pair of nodes strictly along each $E(\mathbf{k}=\Gamma M)$ line, see figure \ref{fig:tilt}(e). Once one deviates from the $\Gamma M$ lines, a finite SOC together with the Zeeman field induces non-unitary equal spin pairing, which opens up a gap around the Fermi level. Therefore, the nodes appear strictly along $\Gamma M$, where no triplets are allowed to exist.

\begin{figure}
\centering
\includegraphics[width=0.35\textwidth]{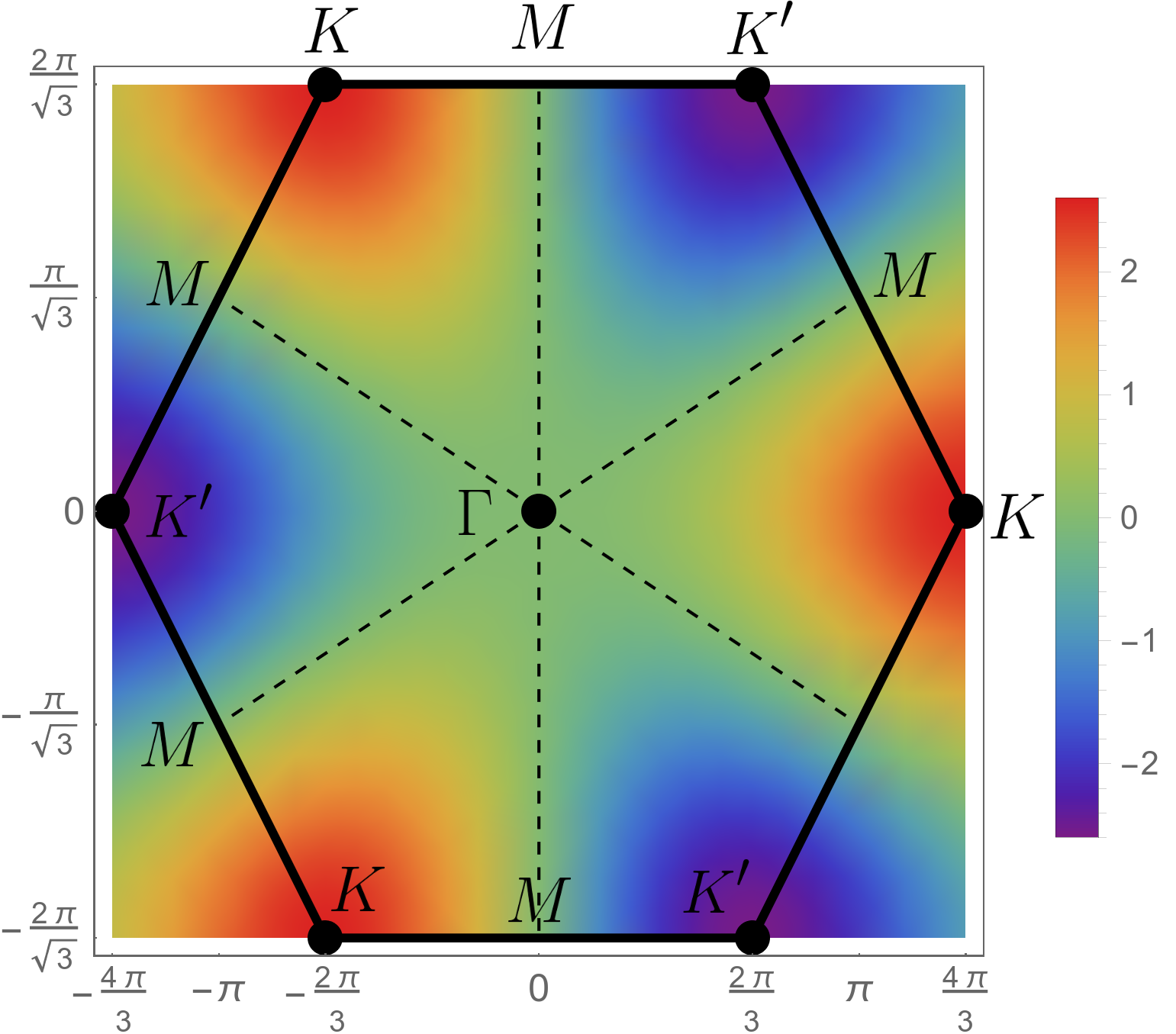}
\caption{\label{fig:bsoc}
Contour plot of $\mathbf{g}(\mathbf{k})\parallel d_z(\mathbf{k})\hat{\mathbf{z}}$ according to equation \eqref{eq:isingSOC} with the first Brillouin zone indicated by the hexagon. The colors show the anti-symmetric sign modulation. The $g$-vector vanishes along the $\Gamma M$ lines, which are indicated by the dashed lines. Since, $\mathbf{g}(\mathbf{k})=0$ along $\Gamma M$, no triplets exist along this line, and superconductivity is unprotected against Zeeman splitting.} 
\end{figure} 

\section{Electron-hole symmetry \label{app:electronhole}}

In this section, we discuss the restrictions that electron-hole symmetry imposes on the anomalous (superconducting) Green's functions.
Bogoliubov-deGennes (BdG) Hamiltonians have built-in electron-hole symmetry, such that the centre of the eigenvalue spectrum is $\mathfrak{b}=0$. 
The discussion here generalizes some remarks made by Berthod \cite{Berthod2016} for the multi-orbital case, and including both singlet and triplet pairing channels. 

A diagonalized BdG Hamiltonian can be written as $H=\sum_\alpha |\alpha\rangle E_\alpha\langle \alpha |$, where $\{E_\alpha\}$ are the eigenvalues and $\{|\alpha\rangle\}$ the eigenvectors of $H$. We can define electron and hole amplitudes as $u^\alpha_{\mathbf{i}\mu\sigma}=\langle \alpha|c^\dag_{\mathbf{i}\mu\sigma}\rangle$ and $v^\alpha_{\mathbf{i}\mu\sigma}=\langle \alpha|c_{\mathbf{i}\mu\sigma}\rangle$ respectively. Due to electron-hole symmetry of the BdG Hamiltonian, if $(E_\alpha,u^\alpha_{\mathbf{i}\mu\sigma},v^\alpha_{\mathbf{i}\mu\sigma})$ is a solution of the Hamiltonian, then $(-E_\alpha,v^{\alpha *}_{\mathbf{i}\mu\sigma},-u^{\alpha *}_{\mathbf{i}\mu\sigma})$ is also a solution \cite{Zhu2016,Sato2017}. For the anomalous Green's function, this implies that 

\begin{equation}
\begin{split}
F_{\mathbf{ij},\sigma\sigma^\prime}^{\mu\nu}(z) & =\langle c_{\mathbf{i}\mu\sigma}|(z-H)^{-1}|c_{\mathbf{j}\nu\sigma^\prime}\rangle\\
& = \sum_\alpha\frac{\langle c_{\mathbf{i}\mu\sigma}|\alpha\rangle\langle\alpha|c_{\mathbf{j}\nu\sigma^\prime}\rangle}{z-E_\alpha}=
\sum_\alpha\frac{u^{\alpha *}_{\mathbf{i}\mu\sigma} v^{\alpha }_{\mathbf{j}\nu\sigma^\prime}}{z-E_\alpha}\\
& = -\sum_\alpha\frac{v^{\alpha }_{\mathbf{i}\mu\sigma}u^{\alpha *}_{\mathbf{j}\nu\sigma^\prime}}{z+E_\alpha}=
\sum_\alpha\frac{\langle\alpha|c_{\mathbf{i}\mu\sigma}\rangle\langle c_{\mathbf{j}\nu\sigma^\prime}|\alpha\rangle}{-z-E_\alpha}\\
&= \langle c_{\mathbf{j}\nu\sigma^\prime}|(-z-H)^{-1}|c_{\mathbf{i}\mu\sigma}\rangle=
F_{\mathbf{ji},\sigma^\prime\sigma}^{\nu\mu}(-z).
\end{split}
\label{eq:eh}
\end{equation}
 
To calculate the anomalous Green's functions $F_{\mathbf{ij},\sigma\sigma^\prime}^{\mu\nu}(z)$ we need the expansion moments $\langle c_{\mathbf{i}\mu\sigma}|T_n(\tilde{H})|c_{\mathbf{j}\nu\sigma^\prime}\rangle$. 
For BdG Hamiltonian with $\mathfrak{b}=0$, the rescaled Hamiltonian $\tilde{H}$ has the same symmetries as $H$. Therefore, analogously to the procedure in equation \eqref{eq:eh}, we have 
\begin{equation}
\begin{split}
\langle c_{\mathbf{i}\mu\sigma}|T_n(\tilde{H})|c_{\mathbf{j}\nu\sigma^\prime}\rangle & = \sum_\alpha u^{\alpha *}_{\mathbf{i}\mu\sigma}T_n(\tilde{E}_\alpha)v^{\alpha}_{\mathbf{j}\nu\sigma^\prime}\\
& = -\sum_\alpha v^{\alpha }_{\mathbf{i}\mu\sigma}T_n(-\tilde{E}_\alpha)u^{\alpha *}_{\mathbf{j}\nu\sigma^\prime}\\
&=(-1)^{n+1}\langle c_{\mathbf{j}\nu\sigma^\prime}|T_n(\tilde{H})|c_{\mathbf{i}\mu\sigma}\rangle,
\end{split}
\label{eq:tneh}
\end{equation}
where in the last step we used the property of the Chebyshev polynomials $T_n(-x)=(-1)^n T_n(x)$.

We can use the symmetries obtained in equations \eqref{eq:eh} and \eqref{eq:tneh} to simplify the difference of the anomalous Green's functions evaluated immediately above and below the real axis $F_{\mathbf{ij},\sigma\sigma^\prime}^{\mu\nu}(E+i0)-F_{\mathbf{ij},\sigma\sigma^\prime}^{\mu\nu}(E-i0)$, which is what is needed to calculate the superconducting order parameters. Therefore, together with the Chebyshev expansion \eqref{eq:expansion} we have

\begin{widetext}
\begin{equation}
\begin{split}
&F_{\mathbf{ij},\sigma\sigma^\prime}^{\mu\nu}(E+i0)-F_{\mathbf{ij},\sigma\sigma^\prime}^{\mu\nu}(E-i0)  = 
F_{\mathbf{ij},\sigma\sigma^\prime}^{\mu\nu}(E+i0)-F_{\mathbf{ji},\sigma^\prime\sigma}^{\nu\mu}(-E+i0)=\\
&  =-\frac{1}{\mathfrak{a}}\frac{i}{\sqrt{1-\tilde{E}^2}}\sum_{n=0}^\infty(2-\delta_{n0})\left[\langle c_{\mathbf{i}\mu\sigma}|T_n(\tilde{H})|c_{\mathbf{j}\nu\sigma^\prime}\rangle e^{-in\arccos\tilde{E}}-\langle c_{\mathbf{j}\nu\sigma^\prime}|T_n(\tilde{H})|c_{\mathbf{i}\mu\sigma}\rangle e^{-in\arccos(-\tilde{E})}\right]\\
&  =-\frac{1}{\mathfrak{a}}\frac{i}{\sqrt{1-\tilde{E}^2}}\sum_{n=0}^\infty(2-\delta_{n0})\langle c_{\mathbf{i}\mu\sigma}|T_n(\tilde{H})|c_{\mathbf{j}\nu\sigma^\prime}\rangle \left[ e^{-in\arccos\tilde{E}}-(-1)^{n+1}e^{-in\arccos(-\tilde{E})}\right]\\
& =-\frac{1}{\mathfrak{a}}\frac{i}{\sqrt{1-\tilde{E}^2}}\sum_{n=0}^\infty(2-\delta_{n0})\langle c_{\mathbf{i}\mu\sigma}|T_n(\tilde{H})|c_{\mathbf{j}\nu\sigma^\prime}\rangle 2\cos\left(n\arccos\tilde{E}\right).
\end{split}
\end{equation}
\end{widetext}
This result is used to obtain \eqref{eq:gapB} from \eqref{eq:gap}.

\section{\label{app:kpm}Computational details}

In this section we comment on two technical details related to the Chebyshev expansion method: Chebyshev-Gauss integration and fast Fourier transforms. 

\subsection{Chebyshev-Gauss quadrature}
Given $N$ discretized Chebyshev points $x_k=\cos[\pi/N(k+1/2)]$, integrals can be approximated by a Chebyshev-Gauss quadrature\cite{Weisse2005b}
\begin{equation}
\begin{split}
\int_{-1}^1\mathrm{d}x\,f(x)g(x)&\approx\frac{1}{N}\sum_{k=0}^{N-1}\pi \sqrt{1-x_k^2} f(x_k)g(x_k) \\
&= \frac{1}{N}\sum_{k=0}^{N-1}\gamma_k g(x_k).
\end{split}
\label{eq:gauss}
\end{equation}
This was used to obtain the final form of the temperature dependent coefficients $D_n$ in equation \eqref{eq:dn}. The advantage of this is that the integrals not only become simple sums, but have the form of a fast Fourier transform, which then allows an efficient evaluation.

\subsection{Fast Fourier transformation}
One can obtain energy discretized spectral quantities efficiently using fast Fourier transforms.
If we discretize the retarded Green's function $G(E+i0)$ on the Chebyshev interval $T_n(\tilde{E}_k)=0\Rightarrow \tilde{E}_k=\cos[\pi/N(k+1/2)]$, then equation \eqref{eq:expansion} gives  
\begin{equation}
\tilde{G}(\tilde{E}_k+i0)=\sum_{n=0}^{N-1}\frac{(2-\delta_{n0})}{i\sqrt{1-\tilde{E}^2_k}}\,\mu_n e^{-i\frac{n\pi}{\tilde{N}}\left(k+\frac{1}{2}\right)}.
\label{eq:directly}
\end{equation}
Here the expansion moments $\mu_n$ are to be understood as some matrix element of interest $\langle\alpha| T_n(\tilde{H})|\beta\rangle$, already recursively obtained. The Green's function \eqref{eq:directly} also carries the indices $\alpha$ and $\beta$, but are omitted for simplicity. Each one of these indices specify the lattice site, the orbital and the spin projection.
The form of the function
\begin{equation}
\Gamma_k  =i\sqrt{1-\tilde{E}^2_k}\,\tilde{G}(\tilde{E}_k+i0)
= \sum_{n=0}^{N-1}(2-\delta_{n0})\mu_n e^{-i\frac{n\pi}{2N}}e^{-i\frac{n\pi}{N}k}
\end{equation}
closely resembles that of a standard fast Fourier transformations of the type (apart from a factor of $2$ in the last exponential)
\begin{equation}
\Lambda_k=\sum_{n=0}^{N-1}\lambda_n e^{-\frac{2\pi i n}{N}k},
\label{eq:FFT}
\end{equation}
for which many numerical libraries are available. 
For even $k=2l$, with $l\in \mathbb{N}$ we have
\begin{equation}
\Gamma_{2l}=
\sum_{n=0}^{N-1}\overbrace{\left(2-\delta_{n0}\right)\mu_n  e^{-i\frac{n\pi}{2N}}}^{\lambda_n} e^{-\frac{2\pi i n}{N} l},
\end{equation}
which is of the same form of the desired Fourier transform \eqref{eq:FFT}. For odd $k=2l+1$ and the algebraic manipulations
\begin{equation}
\begin{split}
& \exp-\frac{in\pi}{2\tilde{N}} \exp -\frac{2\pi i n (\tilde{N}-1-l)}{\tilde{N}}=\\
&\exp -\frac{in\pi}{2\tilde{N}}\left[1+4(\tilde{N}-1-l)\right]=\\
&
\exp -\frac{in\pi}{\tilde{N}}\left[\frac{1}{2}+2\tilde{N}-1-k\right]=\exp i\frac{n\pi}{\tilde{N}}\left(k+\frac{1}{2}\right),  
\end{split}
\end{equation}
where in the third line we made the identification $k\rightarrow 2l+1$,
we can write
\begin{equation}
\Gamma_{2l+1}^*=
\sum_{n=0}^{N-1}\left(2-\delta_{n0}\right)\mu_n^* e^{-i\frac{n\pi}{2N}} e^{-\frac{2\pi i n}{N}(\tilde{N}-1-l)}.
\label{eq:manip}
\end{equation}
We can therefore use equation \eqref{eq:FFT} with
\begin{equation}
\lambda_n=\left(2-\delta_{n0}\right)\bar{\mu}_n e^{-i\frac{n\pi}{2N}},\quad
\bar{\mu}_n = \begin{cases} \mu_n, & \mbox{if } n\mbox{ is even} \\ \mu_n^*, & \mbox{if } n\mbox{ is odd} \end{cases},
\end{equation}
with $k\in [0,\tilde{N}-1]$. We then have the relations $\Gamma_{2l}=\Lambda_l$ and $\Gamma_{2l+1}=\Lambda^*_{\tilde{N}-1-l}$, with $l\in [0,\tilde{N}/2-1]$. If the original Hamiltonian is real, then $\mu_n=\mu_n^*$, and a single fast Fourier transform is sufficient. However, if the Hamiltonian has imaginary elements, then one has to do two fast Fourier transforms, one for the even $\mu_n$, and another for the odd $\mu_n^*$.

The scheme described above is useful if one needs the whole spectral range of the Hamiltonian. If only a few energy points are needed, to resolve the superconducting gap around the Fermi level for instance, then one can just evaluate equation \eqref{eq:directly} on the points $\tilde{E}$ of interest.

\begin{table*}
\centering
\caption{The 18 real-space tight binding matrices $[t_{\mathbf{ij},n}^{\mu\nu}]_{3\times 3}$ in the basis of  $\{d_{z^2},d_{xy},d_{x^2-y^2}\}$. The six rows ($n$) refer to the neighboring directions, and the three columns to the first  $\hat{R}_n$, second  $\hat{S}_n$, and third-nearest neighbor $\hat{T}_n$ respectively. 
For each fixed direction and neighbor, the correspondent $3\times 3$ matrix contains the diagonal intra-orbital hoppings ($\mu=\nu$), and the off-diagonal inter-orbital hoppings ($\mu\neq \nu$).
The $\hat{T}_n$ have the same direction as the $\hat{R}_n$, and  therefore also have the same form. Note the symmetry $t_{\mathbf{ij},n}^{\mu\nu}=t_{\mathbf{ij},n+3}^{\nu\mu}$ and the properties in equations \eqref{eq:sumR} and \eqref{eq:sumS}. }
\label{tab:tbtable}
\begin{tabular}{|c|c|c|c|c|c|c|c|c|c|}
\hline
\multicolumn{1}{|l|}{$n$} & \multicolumn{3}{c|}{$\hat{R}_n$} & \multicolumn{3}{c|}{$\hat{S}_n$} & \multicolumn{3}{c|}{$\hat{T}_n$} \\ \hline
\multirow{3}{*}{$1$} & $t_0$ & $-t_1$ & $t_2$ & $r_0$ & $r_2$ & $-\frac{r_2}{\sqrt{3}}$ & $u_0$ & $-u_1$ & $u_2$ \\ \cline{2-10} 
 & $t_1$ & $t_{11}$ & $-t_{12}$ & $r_1$ & $r_{11}$ & $r_{12}$ & $u_1$ & $u_{11}$ & $-u_{12}$ \\ \cline{2-10} 
 & $t_2$ & $t_{12}$ & $t_{22}$ & $-\frac{r_1}{\sqrt{3}}$ & $r_{12}$ & $r_{11}+\frac{2r_{12}}{\sqrt{3}}$ & $u_2$ & $u_{12}$ & $u_{22}$ \\ \hline
\multirow{3}{*}{$2$} & $t_0$ & $-\frac{t_1+\sqrt{3} t_2}{2}$ & $-\frac{-\sqrt{3}t_1+t_2}{2}$ & $r_0$ & $0$ & $\frac{2r_1}{\sqrt{3}}$ & $u_0$ & $-\frac{u_1+\sqrt{3} u_2}{2}$ & $-\frac{-\sqrt{3}u_1+u_2}{2}$ \\ \cline{2-10} 
 & $\frac{t_1-\sqrt{3}t_2}{2}$ & $\frac{t_{11}+3t_{22}}{4}$ & \begin{tabular}[c]{@{}c@{}}$-\frac{\sqrt{3}}{4}(t_{11}-t_{22})$\\ $+t_{12}$\end{tabular} & $0$ & $r_{11}+\sqrt{3}r_{12}$ & $0$ & $\frac{u_1-\sqrt{3}u_2}{2}$ & $\frac{u_{11}+3u_{22}}{4}$ & \begin{tabular}[c]{@{}c@{}}$-\frac{\sqrt{3}}{4}(u_{11}-u_{22})$\\ $+u_{12}$\end{tabular} \\ \cline{2-10} 
 & $-\frac{\sqrt{3}t_1+t_2}{2}$ & \begin{tabular}[c]{@{}c@{}}$-\frac{\sqrt{3}}{4}(t_{11}-t_{22})$\\ $-t_{12}$\end{tabular} & $\frac{3t_{11}+t_{22}}{4}$ & $\frac{2r_2}{\sqrt{3}}$ & $0$ & $r_{11}-\frac{r_{12}}{\sqrt{3}}$ & $-\frac{\sqrt{3}u_1+u_2}{2}$ & \begin{tabular}[c]{@{}c@{}}$-\frac{\sqrt{3}}{4}(u_{11}-u_{22})$\\ $-u_{12}$\end{tabular} & $\frac{3u_{11}+u_{22}}{4}$ \\ \hline
\multirow{3}{*}{$3$} & $t_0$ & $\frac{t_1+\sqrt{3} t_2}{2}$ & $-\frac{-\sqrt{3}t_1+t_2}{2}$ & $r_0$ & $-r_2$ & $-\frac{r_2}{\sqrt{3}}$ & $u_0$ & $\frac{u_1+\sqrt{3}u_2}{2}$ & $-\frac{-\sqrt{3}u_1+u_2}{2}$ \\ \cline{2-10} 
 & $\frac{-t_1+\sqrt{3}t_2}{2}$ & $\frac{t_{11}+3t_{22}}{4}$ & \begin{tabular}[c]{@{}c@{}}$\frac{\sqrt{3}}{4}(t_{11}-t_{22})$\\ $-t_{12}$\end{tabular} & $-r_1$ & $r_{11}$ & $-r_{12}$ & $\frac{-u_1+\sqrt{3}u_2}{2}$ & $\frac{u_{11}+3u_{22}}{4}$ & \begin{tabular}[c]{@{}c@{}}$\frac{\sqrt{3}}{4}(u_{11}-u_{22})$\\ $-u_{12}$\end{tabular} \\ \cline{2-10} 
 & $-\frac{\sqrt{3}t_1+t_2}{2}$ & \begin{tabular}[c]{@{}c@{}}$\frac{\sqrt{3}}{4}(t_{11}-t_{22})$\\ $+t_{12}$\end{tabular} & $\frac{3t_{11}+t_{22}}{4}$ & $-\frac{r_1}{\sqrt{3}}$ & $-r_{12}$ & $r_{11}+\frac{2r_{12}}{\sqrt{3}}$ & $-\frac{\sqrt{3}u_1+u_2}{2}$ & \begin{tabular}[c]{@{}c@{}}$\frac{\sqrt{3}}{4}(u_{11}-u_{22})$\\ $+u_{12}$\end{tabular} & $\frac{3u_{11}+u_{22}}{4}$ \\ \hline
\multirow{3}{*}{$4$} & $t_0$ & $t_1$ & $t_2$ & $r_0$ & $r_1$ & $-\frac{r_1}{\sqrt{3}}$ & $u_0$ & $u_1$ & $u_2$ \\ \cline{2-10} 
 & $-t_1$ & $t_{11}$ & $t_{12}$ & $r_2$ & $r_{11}$ & $r_{12}$ & $-u_1$ & $-u_{11}$ & $u_{12}$ \\ \cline{2-10} 
 & $t_2$ & $-t_{12}$ & $t_{22}$ & $-\frac{r_2}{\sqrt{3}}$ & $r_{12}$ & $r_{11}+\frac{2r_{12}}{\sqrt{3}}$ & $u_2$ & $-u_{12}$ & $u_{22}$ \\ \hline
\multirow{3}{*}{$5$} & $t_0$ & $\frac{t_1-\sqrt{3}t_2}{2}$ & $-\frac{\sqrt{3}t_1+t_2}{2}$ & $r_0$ & $0$ & $\frac{2r_1}{\sqrt{3}}$ & $u_0$ & $\frac{u_1-\sqrt{3}u_2}{2}$ & $-\frac{\sqrt{3}u_1+u_2}{2}$ \\ \cline{2-10} 
 & $-\frac{t_1+\sqrt{3}t_2}{2}$ & $\frac{t_{11}+3t_{22}}{4}$ & \begin{tabular}[c]{@{}c@{}}$-\frac{\sqrt{3}}{4}(t_{11}-t_{22})$\\ $-t_{12}$\end{tabular} & $0$ & $r_{11}+\sqrt{3}r_{12}$ & $0$ & $-\frac{u_1+\sqrt{3}u_2}{2}$ & $\frac{u_{11}+3u_{22}}{4}$ & \begin{tabular}[c]{@{}c@{}}$-\frac{\sqrt{3}}{4}(u_{11}-u_{22})$\\ $-u_{12}$\end{tabular} \\ \cline{2-10} 
 & $-\frac{-\sqrt{3}t_1+t_2}{2}$ & \begin{tabular}[c]{@{}c@{}}$-\frac{\sqrt{3}}{4}(t_{11}-t_{22})$\\ $+t_{12}$\end{tabular} & $\frac{3t_{11}+t_{22}}{4}$ & $\frac{2r_1}{\sqrt{3}}$ & $0$ & $r_{11}-\frac{r_{12}}{\sqrt{3}}$ & $-\frac{-\sqrt{3}u_1+u_2}{2}$ & \begin{tabular}[c]{@{}c@{}}$-\frac{\sqrt{3}}{4}(u_{11}-u_{22})$\\ $+u_{12}$\end{tabular} & $\frac{3u_{11}+u_{22}}{4}$ \\ \hline
\multirow{3}{*}{$6$} & $t_0$ & $\frac{-t_1+\sqrt{3}t_2}{2}$ & $-\frac{\sqrt{3}t_1+t_2}{2}$ & $r_0$ & $-r_1$ & $-\frac{r_1}{\sqrt{3}}$ & $u_0$ & $\frac{u_1+\sqrt{3}t_2}{2}$ & $-\frac{\sqrt{3}u_1+u_2}{2}$ \\ \cline{2-10} 
 & $\frac{t_1+\sqrt{3}t_2}{2}$ & $\frac{t_{11}+3t_{22}}{4}$ & \begin{tabular}[c]{@{}c@{}}$\frac{\sqrt{3}}{4}(t_{11}-t_{22})$\\ $+t_{12}$\end{tabular} & $-r_2$ & $r_{11}$ & $-r_{12}$ & $\frac{u_1+\sqrt{3}u_2}{2}$ & $\frac{u_{11}+3u_{22}}{4}$ & \begin{tabular}[c]{@{}c@{}}$\frac{\sqrt{3}}{4}(u_{11}-u_{22})$\\ $+u_{12}$\end{tabular} \\ \cline{2-10} 
 & $-\frac{-\sqrt{3}t_1+t_2}{2}$ & \begin{tabular}[c]{@{}c@{}}$\frac{\sqrt{3}}{4}(t_{11}-t_{22})$\\ $-t_{12}$\end{tabular} & $\frac{3t_{11}+t_{22}}{4}$ & $-\frac{r_2}{\sqrt{3}}$ & $-r_{12}$ & $r_{11}+\frac{2r_{12}}{\sqrt{3}}$ & $-\frac{-\sqrt{3}u_1+u_2}{2}$ & \begin{tabular}[c]{@{}c@{}}$\frac{\sqrt{3}}{4}(u_{11}-u_{22})$\\ $-u_{12}$\end{tabular} & $\frac{3u_{11}+u_{22}}{4}$ \\ \hline
\end{tabular}

\end{table*}

\section{Group theory}

\subsection{The structure of the superconducting order parameter at \texorpdfstring{$K$}{K}-point} 

In this section, we construct the most general superconducting order parameter within the three-orbital model consistent with the symmetry.
Our strategy is to construct the irreps of the full space group based on the star of the vector $K$.
The Cooper pairs are then obtained by projecting the $A_1'$-symmetric part of the anti-symmetric squares of these irreps.

To construct the irreps of the space group based on the star of $K$, we follow the standard procedure, and build the irreps of the (double) group of $K$, $C_{3h}'$, the so-called \textit{little group irreps}.
In the spinless case, this procedure has been performed in \cite{Kormanyos2015}, and we extend it here to the case of particles with spin.

The star of $K$ contains two rays, $K$ and $K'$. As the group of $K$ $C_{3h}'$ is abelian, all of its irreps are one-dimensional and as a result, the irreps of the space group based on the star of $K$ are two dimensional. 
Similar to the case of no spin, the double group of $K$ is abelian as expected.
We only consider the $xy$ and $x^2-y^2$ orbitals for clarity.
The four-dimensional space of the Bloch orbitals splits into four one dimensional spaces,
\begin{align}\label{4irr}
\psi_{K,+;\uparrow} & = \frac{1}{\sqrt{N}}\sum_{\mathbf{i}} e^{ i \mathbf{K}\cdot \mathbf{R}_{i} }  c_{\mathbf{i}2+2 \uparrow}
\notag \\
\psi_{K,-;\uparrow} & = \frac{1}{\sqrt{N}}\sum_{\mathbf{i}} e^{ i \mathbf{K}\cdot \mathbf{R}_{i} } c_{\mathbf{i}2-2 \uparrow}
\notag \\
\psi_{K,+;\downarrow} & = \frac{1}{\sqrt{N}}\sum_{\mathbf{i}} e^{ i \mathbf{K}\cdot \mathbf{R}_{i} } c_{\mathbf{i}2+2 \downarrow}
\notag \\
\psi_{K,-;\downarrow} & = \frac{1}{\sqrt{N}}\sum_{\mathbf{i}} e^{ i \mathbf{K}\cdot \mathbf{R}_{i} } c_{\mathbf{i}2-2 \downarrow}\, ,
\end{align}
where  $\mathbf{R}_i$ is the location of the transition metal ion in the unit cell $i$, and $N$ is the number of transition metal atoms.
The vectors $\mathbf{R}_i$ form the triangular Bravais lattice. 
The four Bloch states listed in equation  \eqref{4irr}, transform as $\phantom{}^2\bar{E}_3$, $\phantom{}^1\bar{E}_3$, $\phantom{}^1\bar{E}_3$, and  $ \phantom{}^2\bar{E}_1$ respectively.
The characters of these irreps are listed in table \ref{tab:groupK}.
%
\begin{table*}
\begin{centering}
\caption{Characters of four out of $12$ irreps of the group of $K$, $C_{3h}'$. The listed characters refer to the four irreps realised by the four states in equation \eqref{4irr}. Here $\alpha = \exp(i 2 \pi/3)$.\label{tab:groupK}}
\begin{tabular}{c|c|c|c|c|c|c|c|c|c|c|c|c|}
   $C_{3h}'$       & $E$ &      $C_3$     &  $C_3^2$ & $Q $ & $Q C_3$ & $Q C_3^2$
   & $\sigma_h$ &      $s_3$     &  $s_3^2$ & $Q \sigma_h $ & $Q s_3$ & $Q s_3^2$
    \\
\hline
$\phantom{}^2\bar{E}_3 $ & $1$ & $\alpha^{1/2}$ & $\alpha$& $-1$ & $-\alpha^{1/2}$ & $- \alpha$
& $-i$ & $ - i \alpha^{1/2}$ & $- i \alpha$& $ i $ & $i \alpha^{1/2}$ & $ i\alpha$
\\
$\phantom{}^1\bar{E}_3$ & $1$ & $\alpha^{-1/2}$ & $\alpha^{-1}$& $-1$ & $-\alpha^{-1/2}$ & $- \alpha^{-1}$
& $ i $ & $i \alpha^{-1/2}$ & $ i \alpha^{-1}$& $ -i$ & $-i \alpha^{-1/2}$ & $- i\alpha^{-1}$
\\
$\phantom{}^1\bar{E}_1$ & $1$  & $-1$  & $1$  &  $-1$  & $1$ & $-1$
&
$i$  & $-i$  & $i$  &  $-i$  & $i$ & $-i$
\\
$\phantom{}^2\bar{E}_1 $ & $1$  & $-1$  & $1$  &  $-1$  & $1$ & $-1$
&
$-i$  & $i$  & $-i$  &  $i$  & $-i$ & $i$
\\
\hline
\end{tabular} 
\end{centering}
\end{table*}

Let us construct the whole space group irreps now.
Since the considered crystal structure is symmorphic it is sufficient to fix the matrices corresponding to the point group operations forming the $D_{3h}'$ group.
We follow the standard procedure to generate the four irreps of the space group given the four irreps of the group of $K$, equation \eqref{4irr}.
First, we fix the convention for the partner of each of the states listed in equation \eqref{4irr} forming the $K'$ ray of the star as follows,
\begin{align}\label{li}
\bar{\psi}_{K,+;\uparrow} &=U_2'' \psi_{K,+;\uparrow} =  -\psi_{K',-;\downarrow}\notag \\
\bar{\psi}_{K,-;\downarrow} &= U_2'' \psi_{K,-;\downarrow} =  \psi_{K',+;\uparrow}\notag \\
\bar{\psi}_{K,+;\downarrow} &=U_2'' \psi_{K,+;\downarrow} =  \psi_{K',-;\uparrow}\notag \\
\bar{\psi}_{K,-;\uparrow} &= U_2'' \psi_{K,-;\uparrow} =  -\psi_{K',+;\downarrow}\, .
\end{align}
Instead of the rotation by $\pi$ around the $y$-axis, $U_2''$ any other operation transforming the states at $K$ into states at $K'$ could be used.
Notice that up to the sign the partner states coincide with the action of the time reversal operation on the states at $K$.
This is so because $U_2''$ flips the spin as well as transforms 
$(d_{x^2 - y^2} \pm i d_{xy} )$ into $ (d_{x^2 - y^2} \mp i d_{xy} )$.
It follows that the time reversal operation does not require the doubling of the irreps based on $K$.

Having constructed the irreps of the group of $K$, we are going on and construct the whole space group of $K$.
Let us consider for definiteness the two-dimensional irrep of the space group with the basis, $\{\psi_{K,+\uparrow}, \bar{\psi}_{K,+\uparrow} \}$ the other three irreps can be analysed in a similar way.
Any element of the space group can be written as $g=(t_{\mathbf{R}}|D_g) =t_{\mathbf{R}} D_g $, a product of the proper or improper rotation belonging to $D_{3h}'$ and $t_{\mathbf{R}}$ is the translation by a vector of a Bravais lattice $\mathbf{R}$. The translations are represented by diagonal matrices, because 
$(t_{\mathbf{R}}|E)\psi_{K,+;\uparrow} = \exp(- i \mathbf{K} \cdot\mathbf{R}) \psi_{K,+;\uparrow}$ and $(t_{\mathbf{R}}|E)\bar{\psi}_{K,+;\uparrow} = \exp(- i \mathbf{K}' \cdot\mathbf{R}) \bar{\psi}_{K,+;\uparrow}$.
Now the rotational part is constructed differently for the elements in the group of $K$ and for the rest of the elements.
Consider an element in the group of $K$, $D_g \in C_{3h}'$ such element is represented by a diagonal matrix,
\begin{align}\label{eq:space1}
 (t_{\mathbf{0}}|D_g)_{K+\uparrow} = \begin{bmatrix}
 \phantom{}^2\bar{E}_3[D_g] & 0 \\
 0 & \phantom{}^2\bar{E}_3[(U_2'')^{-1}D_g U_2'']
 \end{bmatrix}\, ,
 \end{align}
where the $\phantom{}^2\bar{E}_3$ irrep realised on the state $\Psi_{K+\uparrow}$ is specified in the table \ref{tab:groupK}.
The rest of the elements of the point group, $D_{3h}'$ belong to the coset $U_2'' C_{3h}'$.
Such that for any $g \not\in C_{3h}'$ the representing matrix reads,
\begin{align}\label{eq:space2}
 (t_{\mathbf{0}}|D_g)_{K+\uparrow} = \begin{bmatrix}
0 &  \phantom{}^2\bar{E}_3[D_g U_2'']  \\
\phantom{}^2\bar{E}_3[(U_2'')^{-1}D_g ] & 0
 \end{bmatrix}\, .
 \end{align}
Equations \eqref{eq:space1} and \eqref{eq:space2} give the two-dimensional irreps of the space group based on the state
$\psi_{K+,\uparrow}$ belonging to the star of $K$.
The elements containing no translations, $(t_{\mathbf{0}}|D_g)$ form a subgroup isomorphic to $D_{3h}'$
Viewed in this way, equations \eqref{eq:space1} and \eqref{eq:space2} form the $\bar{E}_1$, see table \ref{tab:ch_double_D3H}.
The same is true for the  $(t_{\mathbf{0}}|D_g)_{K-\downarrow}$.
Yet the $(t_{\mathbf{R}}|D_g)_{K+\uparrow}$ and $(t_{\mathbf{R}}|D_g)_{K-\downarrow}$ are inequivalent irreps of the whole space group which includes translations.
The remaining two states listed in equation \eqref{li} give rise to another pair of inequivalent irreps, $(t_{\mathbf{R}}|D_g)_{K+\downarrow}$ and $(t_{\mathbf{R}}|D_g)_{K-\uparrow}$.
The two sets of matrices $(t_{\mathbf{0}}|D_g)_{K+\downarrow}$ and $(t_{\mathbf{0}}|D_g)_{K-\uparrow}$ form $\bar{E}_3$ irrep of $D_{3h}'$ each, see table \ref{tab:ch_double_D3H}.

\subsection{Projecting the \texorpdfstring{$s$}{s}-wave Cooper pair states out of anti-symmetric products of irreps of a space group}

In the previous section, we have constructed four space group irreps 
$(t_{\mathbf{R}}|D_g)_{K+\uparrow}$, $(t_{\mathbf{R}}|D_g)_{K-\downarrow}$, $(t_{\mathbf{R}}|D_g)_{K+\downarrow}$ and $(t_{\mathbf{R}}|D_g)_{K-\uparrow}$ corresponding to the four states listed in equation \eqref{li}.
As all of these irreps are inequivalent and the superconductivity is $s$-wave we are allowed to extract the $A'_1$ symmetric Cooper pair combinations form each of the four irreps above.
The projection of the anti-symmetric squares of the  four irreps readily produces the four order parameters,
\begin{align}\label{OP_list1}
\hat{\Psi}_1 &= \psi_{K;+\uparrow} \psi_{K';-\downarrow}  -  \psi_{K';-\downarrow} \psi_{K;+\uparrow} \notag \\
\hat{\Psi}_2 &= \psi_{K;-\uparrow} \psi_{K';+\downarrow}  -  \psi_{K';+\downarrow} \psi_{K;-\uparrow} \notag \\
\hat{\Psi}_3 &= \psi_{K;+\downarrow} \psi_{K';-\uparrow}  -  \psi_{K';-\uparrow} \psi_{K;+\downarrow} \notag \\
\hat{\Psi}_4 &= \psi_{K;-\downarrow} \psi_{K';+\uparrow}  -  \psi_{K';+\uparrow} \psi_{K;-\downarrow}\, . 
\end{align}
For each of the four irreps the number of independent $A_1$ order parameters is 
\begin{align}\label{multiplicity}
\frac{1}{ 24 N}\frac{1}{2} \sum_{D_g \in D'_{3h} }\sum_{\mathbf{i}}  \left[ \chi^2[ (t_{\mathbf{R}_{\mathbf{i}}}| D_g)] -  \chi[ (t_{\mathbf{R}_{\mathbf{i}}}| D_g)^2] \right] =1\, ,
\end{align}
which means that equation \eqref{OP_list1} exhausts all possible Cooper pairs at $K(K')$.
This also shows that there are no cross correlations in the $A_{1}'$ symmetric state,
$ \langle \psi_{\mathbf{K};\pm\uparrow (\downarrow)} \psi_{\mathbf{K}';\mp\uparrow(\downarrow)} \rangle =\langle \psi_{\mathbf{K};\pm\uparrow (\downarrow)} \psi_{\mathbf{K}';\pm\downarrow(\uparrow)} \rangle =0$.

In many cases including monolayer NbSe$_2$, only two band crossings are present near $K(K')$.
As a result, only two out of four combinations listed in equation \eqref{OP_list1} is of practical value,
\begin{equation}
\label{OP_list2}
\hat{\Psi}_1 = \psi_{K;+\uparrow} \psi_{K';-\downarrow},\quad
\hat{\Psi}_4 = \psi_{K;+\downarrow} \psi_{K';-\downarrow}.
\end{equation} 
Out of the two order parameters in equation \eqref{OP_list2}, the singlet and triplet can be formed,
\begin{align}
\label{OP_list3}
\hat{\Psi}_{K,\mathrm{singlet}} &= \frac{1}{2} (\hat{\Psi}_1 + \hat{\Psi}_4 ),\nonumber \\
\hat{\Psi}_{K,\mathrm{triplet}} &= \frac{1}{2} (\hat{\Psi}_1 - \hat{\Psi}_4 ).
\end{align} 
The presence of the finite triplet correlations is due to the lack of inversion symmetry and the SOC.
As the SOC is turned off, the order parameters $\hat{\Psi}_1$ and $\hat{\Psi}_4$ merge into a singlet, 
and the triplet order parameter vanishes accordingly.
The splitting between $\hat{\Psi}_1$ and $\hat{\Psi}_4$ can be traced to the spin splitting of bands at $K(K')$.
No such splitting occurs along $\Gamma M$ and as a result, the superconducting order parameter along $\Gamma M$ is a pure singlet.

\subsection{Bi-linear scalar combinations constructed from operator sets of definite symmetry}
\label{app:invariants}

In this section, we derive all the possible bi-linear combinations out of the operators of a prescribed symmetry.
We denote the symmetry group as $G$ with the number of elements $n_G$.
We imagine having the (not necessarily Hermitian) operators $O_{\kappa;\alpha,j}$ transforming as the a given $\alpha$th representation of dimension $N_{\alpha}$.
The index $j$ enumerates all the functions transforming as $\alpha$, such that $j = 1,\ldots, N_{\alpha}$.
The index $\kappa$ enumerates the multiple sets of operators transforming as the same irrep, $\alpha$.
If there are $m_{\alpha}$ distinct sets of operators transforming as $\alpha$, $\kappa=1,\ldots,m_{\alpha}$.

We aim at listing all the linear combinations of products, $[O_{\kappa;\alpha,j}]^{\dagger} O_{\kappa';\beta,j'}$ that are symmetric under the group $G$. 
By assumption, the operators transform as
\begin{align}\label{grt1}
    [O_{\kappa;\alpha,j}]'=
    \sum_{l=1}^{N_{\alpha}} T^{\alpha}_{ l j}(g) O_{\kappa;\alpha,l}
\end{align}
Without loss of generality,
the matrices can be assumed to be independent of $\kappa_{\alpha}$ for a given $\alpha$ and unitary, $[T^{\alpha}(g)]^{\dagger} =[T^{\alpha}(g)]^{-1}$.
By taking the Hermitian conjugation of equation \eqref{grt1} we obtain the transformation law
\begin{align}\label{grt2}
    \left[O^{\dagger}_{\kappa;\alpha,j}\right]' =
    \sum_{l=1}^{N_{\alpha}} [T^{\alpha}_{ l j}(g)]^* O^{\dagger}_{\kappa;\alpha,l}\, .
\end{align}
The last equation signifies the Hermitian conjugates, $O^{\dagger}_{\kappa_{\alpha};\alpha,l}$ as transforming according to the conjugated irrep.
These irreps may or may not be equivalent to the original irreps in the cases when the characters are real or not.
This is immaterial to the statement we are about to show.

Consider the projection $P_A$ of any of the products, $[O_{\kappa_{\alpha};\alpha,j}]^{\dagger} O_{\kappa_{\beta};\beta,j'}$ onto the trivial irrep, we call $A$.
As all the characters of $A$ are equal to unity,
\begin{align}
P_A[  O^{\dagger}_{\kappa;\alpha,j} O_{\kappa';\beta,j'} ] =& \frac{1}{n_G} \sum_{g \in G} \sum_{l=1}^{N_{\alpha}}\sum_{k=1}^{N_{\beta}}
[T^{\alpha}_{l j}(g)]^* T^{\beta}_{k j'}(g)  
\notag \\
& \times O^{\dagger}_{\kappa;\alpha,l} O_{\kappa';\beta,k}.
\end{align}
Or, using the orthogonality properties of the matrices of unitary irreps, we obtain
\begin{align}
P_A & [  O^{\dagger}_{\kappa;\alpha,j} O_{\kappa';\beta,j'} ] =
\delta_{\alpha\beta}\delta_{jj'}  \sum_{l=1}^{N_{\alpha}}\sum_{k=1}^{N_{\beta}}
 \delta_{lk}
 O^{\dagger}_{\kappa;\alpha,l} O_{\kappa';\beta,k}
 \notag \\
 = &
\delta_{\alpha\beta}\delta_{jj'}  \sum_{l=1}^{N_{\alpha}}
 O^{\dagger}_{\kappa;\alpha,l} O_{\kappa';\alpha,l}.
\end{align}
Therefore the most general scalar, $S$ of a kind considered reads
\begin{align}
S = \sum_{\alpha;\kappa,\kappa'} U_{\alpha;\kappa,\kappa'} \left[\sum_{l=1}^{N_{\alpha}}
 O^{\dagger}_{\kappa;\alpha,l} O_{\kappa';\alpha,l}\right]\, .
\end{align}
In the last sum there are $\sum_{\alpha} m_{\alpha}^2 $ independent combinations with arbitrary {\it complex} amplitudes, $U_{\alpha;\kappa,\kappa'}$.
In the cases when the operator $S$ is required to be Hermitian for each $\alpha$ there are $m_{\alpha} (m_{\alpha} +1)/2$ combinations $ \sum_{l=1}^{N_{\alpha}}
(  O^{\dagger}_{\kappa;\alpha,l} O_{\kappa';\alpha,l}  + O^{\dagger}_{\kappa';\alpha,l} O_{\kappa;\alpha,l}) $ and $m_{\alpha} (m_{\alpha} -1)/2$ combinations $ \sum_{l=1}^{N_{\alpha}}
i (  O^{\dagger}_{\kappa;\alpha,l} O_{\kappa';\alpha,l}  - O^{\dagger}_{\kappa';\alpha,l} O_{\kappa;\alpha,l} )$ with {\it real} coefficients.

For the case when $O$-operators describe the superconducting correlations such as those listed in section~\ref{sec:on-site1}, the combinations of the type $OO$ are not allowed by the particle conservation, and the combinations of  $O^{\dagger}O$ are the only choice.
In the case of Hermitian operators, the irreps are real and the conclusion remains the same.
The combinations belonging to the same symmetry together produce a scalar as described above.

\newpage  

\bibliographystyle{apsrev4-1}
\bibliography{bibliography}

\end{document}